\documentclass[acmsmall,screen,nonacm]{acmart}

\AtBeginDocument{%
  \providecommand\BibTeX{{%
    \normalfont B\kern-0.5em{\scshape i\kern-0.25em b}\kern-0.8em\TeX}}}

\setcopyright{acmcopyright}
\copyrightyear{2022}
\acmYear{2022}
\acmDOI{10.1145/1122445.1122456}

\acmJournal{JEA}

\usepackage{hyperref}

\usepackage{amsmath,amsfonts}
\usepackage{graphicx}
\usepackage{textcomp}
\usepackage{caption}
\usepackage{subcaption}
\usepackage{cleveref}

\usepackage[ruled,vlined,linesnumbered,algo2e]{algorithm2e}

\DontPrintSemicolon
\SetKwComment{kc}{\(\triangleright\)~}{}
\SetKwComment{tcp}{\(\triangleright\) }{}
\SetVlineSkip{0cm}
\SetKwInOut{Output}{Output}
\SetKwInOut{Input}{Input}

\SetCommentSty{mycommfont}

\usepackage{xargs,xspace}
\usepackage[colorinlistoftodos,prependcaption,textsize=tiny,disable]{todonotes}

\newcommandx{\uvc}[2][1=]{\todo[color=magenta!50,#1]{\sf \textbf{\"Umit:} #2}\xspace}

\newcommandx{\ay}[2][1=]{\todo[color=blue!50,#1]{\sf \textbf{Apo:}#2}\xspace}

\newcommandx{\mfb}[2][1=]{\todo[color=green!50,#1]{\sf \textbf{Fatih:} #2}\xspace}

\newcommandx{\jing}[2][1=]{\todo[color=SkyBlue!50,#1]{\sf \textbf{Jing:} #2}\xspace}

\newcommandx{\sgo}{\ensuremath{\text {SGORP}}\xspace}
\newcommandx{\sgos}{\ensuremath{\text {SGO}}\xspace}
\newcommandx{\sgotwdr}{\ensuremath{\text {\sgos-2DR}}\xspace}
\newcommandx{\sgotwds}{\ensuremath{\text {\sgos-2DS}}\xspace}
\newcommandx{\sgothdr}{\ensuremath{\text {\sgos-3DR}}\xspace}
\newcommandx{\sgothds}{\ensuremath{\text {\sgos-3DS}}\xspace}

\begin{document}

\title{\sgo : A Subgradient-based Method for d-Dimensional Rectilinear
    Partitioning}


\author{Muhammed Fat{\.i}h Balin}
\email{balin@gatech.edu}

\author{Xiaojing An}
\email{anxiaojing@gatech.edu}

\author{Abdurrahman Ya\c{s}ar}
\email{ayasar@gatech.edu}

\author{\"{U}mit V. \c{C}ataly\"{u}rek}
\email{umit@gatech.edu}

\authornote{Also with Amazon Web Services. This publication describes work performed at the Georgia Institute of Technology and is not associated with Amazon.}

\affiliation{%
    \institution{School of Computational Science and Engineering, Georgia Institute
        of Technology}
    \city{Atlanta}
    \state{GA}
    \country{USA}
}


\begin{abstract}
    Partitioning for load balancing is a crucial first step to parallelize any type
    of computation. In this work, we propose \sgo, a new spatial partitioning method based
    on {\em Subgradient Optimization}, to solve the $d$-dimensional Rectilinear Partitioning
    Problem (RPP). Our proposed method allows the use of customizable objective functions as
    well as some user-specific constraints, such as symmetric partitioning on
    selected dimensions. Extensive experimental evaluation using
    over 600 test matrices shows that our
    algorithm achieves favorable performance against the state-of-the-art RPP and
    Symmetric RPP algorithms. Additionally, we show the effectiveness of our
    algorithm to do application-specific load balancing using two
    applications as motivation: Triangle Counting and Sparse Matrix Multiplication
    (SpGEMM), where we model their load-balancing problems as $3$-dimensional RPPs.
\end{abstract}



\maketitle

\section{Introduction}

Parallel computing systems have become ubiquitous, and ever-increasing data
necessitates their use. Mapping the data and the computation onto the processors
of these parallel systems is usually the initial parallelization step. However,
a significant amount of those data are irregular, i.e., they do not have easily
observable patterns among their elements. Therefore, efficient partitioning of
the data and computation for mapping is
a difficult problem. In some applications, computational dependency and data can
be represented as graphs/hypergraphs by defining
interactions among entities. Connectivity-based
methods~\cite{berman87,Karypis98-SISC,Catalyurek99,Hendrickson00-PARCO}
can be used to partition those type of irregular data.
Another large class of irregular data comes from applications that deals with
entities in multi-dimensional spaces, such as 3D space and time~\cite{GUO2011}. In many cases,
like n-body problems~\cite{Warren1993,Plimpton03-CPC,Karimabadi06-NMSPF}, computational dependencies
depends on spatial relations of the entities, hence it would be preferred to
partition the data (and work) by keeping the entities that are close-by in space
together. Spatial partitioning achieves that by taking the d-dimensional spatial coordinates of the
input and dividing the space into multi-dimensional rectangles and minimizing the
maximum load (hence minimizing the load imbalance)
\cite{Berger87-TC, Manne96-IWAPC, Pilkington96-TPDS, Ujaldon96-ICS,Saule12-JPDC-spart}.
In this work, we tackle
the generalized spatial partitioning problem of irregular data for parallel processing.


Many different spatial partitioning strategies have been proposed based on various
structural constraints, balancing between flexibility and communication patterns
\cite{Grigni96-IWPAISP,Manne96-IWAPC,Ujaldon96-ICS,Berger87-TC,Saule12-JPDC-spart}.
Among these methods, rectilinear partitioning (a.k.a., generalized block distribution)
\cite{Grigni96-IWPAISP,Manne96-IWAPC} partitions two-dimensional space using straight
lines parallel to each dimension. It is one of the most widely used techniques due
to its simplicity and resulting well-structured communication pattern.

The regularity of the partitioned space makes rectilinear partitioning ideal for many
applications. Limiting the number of neighbors restricts communication into logical
rows or columns of virtual mesh topologies and is beneficial for communication patterns,
simplifying the communication and reasoning of many computational
kernels. For instance, matrix/tensor kernels, such as multiplication, can be naturally
represented using rectilinear partitioning. These properties make rectilinear
partitioning a more attractive choice than other types of spatial partitioning techniques.

The optimal one-dimensional Rectilinear Partitioning Problem (RPP) has a polynomial-time
algorithm~\cite{Nicol94-JPDC}. However, multidimensional RPP is
NP-hard~\cite{Grigni96-IWPAISP}. Very
recently,~\cite{Yasar21-JEA} shows that the symmetric
variant of RPP is also NP-hard. However, in the multidimensional case, the conditional RPP,
which finds the optimal partitioning when partitions of all but one dimension are fixed, is
polynomial-time solvable~\cite{Nicol94-JPDC}.

In this paper, we propose \sgo, a Subgradient Optimization~\cite{boyd2003subgradient}
based method to tackle the
more general, multi-dimensional RPP. A multi-dimensional rectilinear partition of a given $d$-dimensional domain arises
when each of the parts are rectangular volumes whose dimensions exactly match
that of each neighbor at each face. Finding a rectilinear partitioning of a
given $d$-dimensional domain is useful not only in applications where the data
is already in a $d$-dimensional space but also in applications where the
computation can be represented in a $d$-dimensional space. Consider Matrix
Multiplication as an example. Here, the input data lies in a $2$-dimensional
space yet the computation is best represented in a $3$-dimensional space.
Thus, there is a need for methods tackling the RPP in higher dimensions.

We demonstrate our method is efficient in finding high-quality solutions for
multiple variants of rectilinear partitioning. One important property of our
proposed method is that it allows the use of customizable objective functions,
such as minimizing the sum of loads of combinations of tiles with arbitrary
partition sizes for each dimension.

The main contributions of this work are:
\begin{itemize}
    \item Presenting a formulation for the Rectilinear Partitioning Problem (RPP) for the continuous domain (Section
          \ref{Preliminaries}),

    \item Proposing and implementing an efficient iterative method for the RPP
          that generalizes to an arbitrary $d$ dimensions and can also solve the Symmetric
          RPP (SRPP) via constraints (Section~\ref{sec:methodology}),

    \item Demonstrating the superiority of the proposed method over the existing
          state-of-the-art, applying an extensive experimental evaluation on more
          than 600 real-world matrices (Section~\ref{subsec:experimental-load}),

    \item Demonstrating the effectiveness of different customizations of \sgo in $2$-dimensional
          (RPP and SRPP), and $3$-dimensional (Triangle Counting and
          Generalized Sparse Matrix-Matrix Multiplication (SpGEMM)) use-cases
          (Section~\ref{subsec:experimental-load}).
\end{itemize}

In the following sections, we first present the related work for
rectilinear partitioning as well as application use cases in
Section~\ref{sec:relatedwork}.  Next, we present the preliminaries and our
formulation of RPP in Section~\ref{Preliminaries}. Then in
Section~\ref{sec:methodology}, we present the \sgo algorithm.
Section~\ref{sec:mapping} demonstrates how to use \sgo to solve different
variants possibly motivated by real-world use-cases. Finally,
Section~\ref{sec:experiments} presents the detailed experimental
evaluation, and in Section~\ref{sec:conclusion} we conclude.

\section{Related Work}
\label{sec:relatedwork}






\subsection{Rectilinear Partitioning}

Rectilinear partitioning (a.k.a., generalized block distribution) is a well
studied problem~\cite{Nicol94-JPDC,Manne96-IWAPC,Khanna97-ICALP,Aspvall01-TCS,gaur2002constant}.
Two of the existing important algorithms that we cover in the context of this paper for
the (symmetric) rectilinear partitioning problem are Nicol's algorithm~\cite{Nicol94-JPDC} for
$2$-dimensional RPP and the PAL algorithm~\cite{Yasar21-JEA} for the
$2$-dimensional SRPP. The biggest difference between these methods and \sgo is
that \sgo can work in an arbitrary number of dimensions and contains the
$2$-dimensional RPP and SRPP as special cases, while the other algorithms only
work for their individual cases. Nicol's algorithm is an iterative method and uses the fact
that given the partition of a dimension, it is possible to find the optimal
partition for the other dimension. Compared to the Nicol's algorithm, our method
changes the partitions for all the dimensions at the same time in a single step.
On the other hand, PAL algorithm is a single shot heuristic. It makes a single
pass over the matrix nonzeros and partitions the matrix along the way. While the
output partitions for both Nicol's algorithm and PAL algorithm are at local
optima, which was defined in~\eqref{optimality}, we will observe that \sgo will
output partitions that are usually at better local optima.

Aspvall et al.~\cite{Aspvall01-TCS} show that in the existence of
heavy rows/columns, Nicol's algorithm~\cite{Nicol94-JPDC} focuses on
heavy rows/columns, and that causes accumulation of the load in other rows.
To overcome this problem, Aspvall et al.~\cite{Aspvall01-TCS} propose an objective function which ignores the heavy rows/columns and solves the problem by iterating only 2-3 times.

Khanna et al.~\cite{Khanna97-ICALP} and Gaur et al.~\cite{gaur2002constant} propose
mapping rectilinear partitioning problem to the rectangle stabbing problem.
The main drawback of this approach is its rather high computational complexity.
In this process, the first stage involves finding all the rectangles that have higher load than
a given target load, and it can take up to $O(n^4)$ for $n \times n$ dense matrices and
$O(o^2)$ for sparse matrices with $o$ nonzeros. Furthermore, rectangle stabbing algorithms
running on these rectangles have a long runtime. Thus, we don't compare \sgo with this class of algorithms.


There exists an iterative 4-approximation algorithm~\cite{muthu2005}. In this approach,
the authors maintain costs for each individual row and column initialized to 1 at the start.
After, at each step, they find a tile exceeding a given target load and scale the row and column
costs of that tile by $1 + \frac{\epsilon}{2}$, where $\epsilon > 0$. Then,
they partition the row and column cost arrays using the approach we describe in~\ref{subsec:sgo1d}.
One main difference between this algorithm and \sgo is that
their algorithm requires a target load $L^t$ as an input to see whether a partition exists whose
maximum load is less than $L^t$, thus being a solution to a decision problem, whereas \sgo
outputs the best partition found at the end. The other difference is that \sgo utilizes
information that comes from every part of the load matrix and also their magnitudes, however
its counterpart only utilizes the information about the location of the maximally loaded part.

\subsection{Triangle Counting}

The triangle counting problem~\cite{Latapy08-TCS,Hu18-SC,Yasar22-TPDS}
seeks to find mutually connected 3-vertices in an undirected graph.
This is a crucial graph kernel that serves as a building block for many other
graph problems.
For the interest of this paper, recently, Hu et al.~\cite{Hu18-SC} proposed to use
rectilinear partitioning to divide the computation among GPUs and Ya\c{s}ar et
al.~\cite{Yasar22-TPDS} proposed a block-based triangle counting formulation
using symmetric rectilinear partitioning to make the algorithm suitable for task-based
execution on shared and distributed-memory systems. Both approaches try to minimize
the maximum load of a partition. In this work, we show that \sgo can model
this partitioning problem in a three dimensional space and optimize a different
objective function successfully (see Section~\ref{sym_tri}).

\subsection{SUMMA-SpGEMM}

SpGEMM, sparse matrix matrix multiplication, computes matrix multiplication on two sparse
matrices. It is commonly used in graph applications, such as link
prediction~\cite{sarkar2011theoretical, martinez2016survey}, graph compression\cite{navlakha2008graph},
and used in scientific computations~\cite{ravasz2002hierarchical, lin2014towards}.
Due to its high complexity, and extreme irregularity, there has been an interest in
optimizing SpGEMM~\cite{bulucc2012parallel, liu2014efficient, deveci2018multithreaded,
    demirci2020cartesian, lee2020optimization} in both shared memory and distributed systems.
SUMMA-SpGEMM\cite{bulucc2012parallel}, inspired by the original dense
SUMMA~\cite{Van97-CPE}, is one of the most commonly used technique for distributed memory systems.
In SUMMA, the result matrix's computation is partitioned rectilinearly, and each processor
in a 2D virtual processor grid calculates a part. By iteratively generating partial results, space usage for
each process is limited to a constant number of parts of matrices.
We show that \sgo can directly partition both input matrices simultaneously
while incorporating minimization of the communication volume into partitioning objective (see Section~\ref{nsym_spgemm}).

\section{Preliminaries}\label{Preliminaries}

\subsection{Definitions}

Rectilinear partitioning in the $d$-dimensional space consists of $d$
$1$-dimensional partitions, one for each of the $d$-dimensions. That's why we
will start by defining what it means to partition a $1$-dimensional interval.

Given an interval $r = [a, b)$, a partitioning $p$ of $r$ into $k$ parts is an
array $[p[0], \dots, p[k]]$ such that $p[0] = a$, $p[k] = b$ and it is
monotonic, i.e., $p[j] \leq p[j + 1], \forall j \in [k]$. Here, we use $[k]$
to represent $\{0, \dots, k - 1\}$ and will use $[k]^+$ to represent $\{1,
    \dots, k\}$.

Next, we will define the objects that we will try to partition. Let us define a
load distribution as an integrable function $f: \mathbb{R}^d \to \mathbb{R}^+$, and let us define its load $L(f)$ as:
\begin{align*}
    L(f) = \int_{-\infty}^\infty \dots \int_{-\infty}^\infty f(x_1, \dots, x_d) \,dx_d \dots \,dx_1 < \infty.
\end{align*}

Given a $d$-dimensional load distribution $f$, we would like to find $p = (p_1, \dots, p_d)$
such that each $p_i$ is a partitioning of $\mathbb{R}$ into $k_i$ parts. Together, the $p_i$'s
imply a partitioning of the whole space $\mathbb{R}^d$ into $k = (k_1, \dots, k_d)$ parts. Our
goal is to minimize $L(f, p)$, the maximum of loads of these parts, i.e.,
\begin{align}
    L^*        & = \min_{p=(p_1, \dots, p_d)} L(f, p)\label{RPP_problem}                                                       \\
    L(f, p)    & = \max_{j \in [k_1] \times \dots \times [k_d]} L(f, p, j)                                                     \\
    L(f, p, j) & = \int_{p_1[j_1]}^{p_1[j_1 + 1]} \dots \int_{p_d[j_d]}^{p_d[j_d + 1]}  f(x_1, \dots, x_d) \,dx_d \dots \,dx_1
    \label{load_definition}
\end{align}

Note that, we seek a $p=(p_1, \dots, p_d)$ to minimize $L(f, p)$, the
load of the maximally loaded part. And finally, let us also define the prefix
sum $F: \mathbb{R}^d \to R^+$ of $f$ as
$$ 
    F(x_1, \dots, x_d) = L(f, [-\infty, x_1], \dots, [-\infty, x_d])
$$ 
and the prefix sum in the $i$-th dimension $F_i: \mathbb{R} \to \mathbb{R}^+$ as
$$ 
    F_i(x) = F(x_1, \dots, x_d) : x_i = x\  \textrm{and}\  x_j = \infty, \forall j \neq i.
$$ 

\begin{table}[ht]
    \small
    \centering
    \caption{Notation used in this paper.}\label{tab:notation}
    \label{tab:definitions}
    \begin{tabular}{r  l}
        \textbf{Symbol}               & \textbf{Description}                                   \\
        \hline
        $[k]$                         & Integer set: $\{0, \dots, k - 1\}$                     \\
        $[k]^+$                       & Integer set: $\{1, \dots, k\}$                         \\
        $[a, b)$                      & Real interval: $\{ x \in \mathbb{R} : a \leq x < b \}$ \\
        $p = (p_1, \dots, p_d)$       & A partitioning of $\mathbb{R}^d$, where $p_i$ is the   \\
                                      & partition array in $i$-th dimension                    \\ 
        $k = (k_1, \dots, k_d)$       & Nu. of parts in each dimension of $p$                  \\
        $L(f, p, j)$                  & Load at index $j = (j_1, \dots, j_d)$ with $p$         \\
        $L(f, p)$                     & Maximum load                                           \\
        $F_i(x)$                      & Prefix sum: $i$-th dimension, point $x$                \\
        $F_i^{-1}(x)$                 & Inverse of $F_i$ at point $x$                          \\
        $A$                           & A sparse tensor                                        \\
        $f_A$                         & Load distribution of the tensor $A$                    \\
        $\pi = (\pi_1, \dots, \pi_d)$ & Parametrization of $p = (p_1, \dots, p_d)$             \\
        $g = (g_1, \dots, g_d)$       & Subgradient at the current parameters                  \\
        $\eta(t)$                     & Step size depending on iteration $t$                   \\
        \hline
    \end{tabular}
\end{table}

Let us denote the inverse of the prefix sum in the $i$-th dimension as
$F_i^{-1}$. We will use these prefix sums in the next section to reparametrize
partitions of each of the $d$ dimensions and it will serve as the basic building
block of our method. A summary of used notation can be found in Table~\ref{tab:definitions}.

\subsection{Modeling sparse tensors and point datasets as load distributions} \label{spmtx}

Let A be a $d$-dimensional sparse tensor with $o$ nonzeros defined via an
ordered index set, $A_I \in \mathbb{N}^{o \times d}$, and corresponding values
of these indices, $A_V \in \mathbb{R}^o$. Thus:
\begin{align*}
    A[A_I[i]] & = A_V[i], \forall i                           \\
    A[e]      & = 0, \forall e \in \mathbb{N}^d \setminus A_I
\end{align*}
Note that, a sparse tensor $A$ is essentially a function from the index space
$\mathbb{N}^d$ to the value space $\mathbb{R}$, i.e., $A: \mathbb{N}^{d} \to
    \mathbb{R}$, and it is zero for any index not in the index set.

In the case of a given $d$-dimensional dataset with $o$ points, we can treat
$A_I \in \mathbb{R}^{o \times d}$ as the coordinates of the points and
$A_V \in \mathbb{R}^o$ as their weights. As there isn't a fundamental difference
between sparse tensors and point datasets in this sense, we will talk about
datasets as sparse tensors.

Given a $d$-dimensional sparse tensor $A$, we will define the corresponding
$d$-dimensional load distribution $f_A$ as follows:
\begin{equation}
    f_A(x_1, \dots, x_d) = \sum_{u \in A_I} \delta(x - u) A[u],
    \label{sparse_tensor}
\end{equation}
where $\delta: \mathbb{R}^d \to \mathbb{R}$ denotes the Dirac delta function to
model a point-wise load that a nonzero in $A$ implies. Defined this way, $f$
represents the distribution of the nonzeros of $A$ in $\mathbb{R}^d$. In the
scenario where the values of the nonzeros of $A$ don't matter, we assume that
all the values are set to $1$.

\section{\sgo: Subgradient Optimization for Rectilinear Partitioning}\label{sec:methodology}

In this section, we explain our customizable framework, \sgo. The reason we are
classifying our method under Subgradient Optimization~\cite{shor85} is that;
we don't have access to the gradient of
our objective function. We pick a direction to move our parameters that will
probably improve the objective but might not every iteration. Given a
$d$-dimensional load distribution $f$, \sgo partitions the load distribution, $f$,
while also taking any user given equality constraints on the partition vectors
of different dimensions to solve the symmetric RPP.
In the following subsections,
we first explain how the algorithm works for the $1$-dimensional partitioning problem.
Afterwards, we show how to generalize our approach to the multi-dimensional
case. Then, we show how we can incorporate equality constraints on the partition
vectors of different dimensions into our framework and how to initialize the
optimization variables. Since \sgo is an iterative method, we also discuss
possible stopping conditions and step size rules. Finally, we give an overall summary
of our method and illustrate a single iteration on a toy example.

\subsection{1-dimensional partitioning problem}\label{subsec:sgo1d}

We first consider the 1-dimensional partitioning problem to build up an
intuition for the RPP. Let $f$ be a $1$-dimensional load distribution.
Our goal is to find a way, $p = (p_1)$, to
partition it into $k = (k_1)$ parts while minimizing $L(f, p)$. Given our
objective, there exists a partition $p^* = (p_1^*)$ such that $L(f, p^*) =
    \frac{L(f)}{k_1}$. Note that $F$ is differentiable by definition, and so it is
also continuous. We can also explicitly express the entries of $p_1^*$ as
follows:
$$
    p_1^*[j] = F^{-1}_1\Big(j\frac{L(f)}{k_1}\Big), \forall j \in [k_1 + 1]
$$
Unfortunately, we can not determine the optimal solution $p^*$ explicitly using
this formulation for higher-dimensional problems. Therefore for a given $p$, we
seek a way to compute its subgradient to improve it iteratively. As one can
notice, $F^{-1}_1$ plays a crucial role to solve the 1-dimensional
partitioning problem. Thus, we argue that in the multidimensional case, it might
also be beneficial to parametrize $p$ with $\pi = (\pi_1, \dots, \pi_d)$ as
follows:
$$ 
    p_i[j] = F^{-1}_i(\pi_i[j])
$$ 
In this parametrization, the optimal solution for the 1-dimensional case
$\pi^* = (\pi_1^*)$ can be expressed as:
$$ 
    \pi_1^*[j] = j \frac{L(f)}{k_1}
$$ 
Then we define the load, $L_\pi$, as:
\begin{align}
    L_\pi(f, \pi)    & = \max_{j \in [k_1] \times \dots \times [k_d]} L_\pi(f, \pi, j) \\
    L_\pi(f, \pi, j) & = L(f, (F_1^{-1}(\pi_1), \dots, F_d^{-1}(\pi_d)), j)
\end{align}
Finally, the subgradient $g = (g_1)$ for the 1-dimensional case can be written as:
$$ 
    g_1[j] = \frac{\partial L_\pi(f, \pi)}{\partial \pi_1} = \pi_1[j] - \pi_1^*[j]
$$ 
and the update rule with step size $\eta(t) > 0$ at iteration $t$ as:
\begin{align}
    \pi' = \pi - \eta(t) g
    \label{update_rule_1d}
\end{align}
Note that when $\eta = 1$, we achieve the optimal solution in one step.

\subsection{Multidimensional partitioning problem}\label{subsec:sgond}

Let $f$ be a $d$-dimensional load distribution and say we want to partition it
into $k = (k_1, \dots, k_d)$ parts. We aim to find a good way to define the
subgradient $g = (g_1, \dots, g_d)$ so that applying the update rule in~\eqref{update_rule_1d} repeatedly, we will get closer and closer to local
optima. But before that, let us define $r_i(f, \pi)[j_i]$ as the maximum over all
dimensions except the $i$th one as follows:
\begin{align}
    r_i(f, \pi)[j_i] = \max_{j \in [k_1] \times \dots \times \{j_i\} \times \dots \times [k_d]} L_\pi(f, \pi, j), \forall j_i \in [k_i]
    \label{ri_computation}
\end{align}

We claim that a given $\pi$ is at local optima in the sense that changing any of
the $\pi_i$ while keeping others fixed will increase the value of $L(f, \pi)$
when the following holds:
$$ 
    r_i(f, \pi)[j_i] = L_\pi(f, \pi), \forall i, j_i
$$ 
Thus, the optimal solution $\pi^*$ lies in the set
\begin{align}
    S = \{\pi \mid r_i(f, \pi)[j_i] = L_\pi(f, \pi), \forall i, j_i\}
    \label{optimality}
\end{align}

In the 1-dimensional case, this set only has a single member, and it is the
optimal solution. However, in the multidimensional case, this set is not
necessarily a singleton.

We will define the subgradient $g = (g_1, \dots, g_d)$ as follows:
\begin{align}
    g_i[j_i] = \sum_{u = 0}^{j_i - 1} r_i[u] - \frac{j_i}{k_i} \sum_{u = 0}^{k_i - 1} r_i[u], \forall i, j_i
\end{align}
so that a given $\pi$ will get closer and closer to the set $S$ by applying our
update rule repeatedly. Let's verify if the subgradient $g$ becomes $0$ when $\pi \in S$:
\begin{align*}
    g_i[j_i] & = \sum_{u = 0}^{j_i - 1} r_i[u] - \frac{j_i}{k_i} \sum_{u = 0}^{k_i - 1} r_i[u]               \\
             & = \sum_{u = 0}^{j_i - 1} L_\pi(f, \pi) - \frac{j_i}{k_i} \sum_{u = 0}^{k_i - 1} L_\pi(f, \pi) \\
             & = j_i L_\pi(f, \pi) - j_i L_\pi(f, \pi) = 0
\end{align*}
Indeed, $g$ becomes $0$ when $\pi$ is at a local optima as expected.

Note that, our update rule in~\eqref{update_rule_1d} reads the same for each of the $d$ dimensions:
\begin{align}
    \pi_i' = \pi_i - \eta_i(t) g_i, \forall i
    \label{update_rule}
\end{align}
Again to verify, in the 1-dimensional case, we have:
\begingroup
\allowdisplaybreaks
\begin{align*}
    \pi'_1[j_1] & = \pi_1[j_1] - \eta g_1[j_1]                                                                           \\
                & = \pi_1[j_1] - \eta \sum_{u = 0}^{j_1 - 1} r_1[u] + \eta \frac{j_1}{k_1} \sum_{u = 0}^{k_1 - 1} r_1[u] \\
                & = \pi_1[j_1] - \eta \sum_{u = 0}^{j_1 - 1} (\pi_1[u + 1] - \pi_1[u])                                   \\
                & \ \ \ + \eta \frac{j_1}{k_1} \sum_{u = 0}^{k_1 - 1} (\pi_1[u + 1] - \pi_1[u])                          \\
                & = \pi_1[j_1] - \eta (\pi_1[j_1] - \pi_1[0]) + \eta \frac{j_1}{k_1} (\pi_1[k_1] - \pi_1[0])             \\
                & = \pi_1[j_1] - \eta (\pi_1[j_1] - 0) + \eta \frac{j_1}{k_1} (L(f) - 0)                                 \\
                & = \pi_1[j_1] - \eta (\pi_1[j_1] - \frac{j_1}{k_1} L(f))                                                \\
                & = \pi_1[j_1] - \eta (\pi_1[j_1] - \pi_1^*[j_1])
\end{align*}
\endgroup

\subsection{Constrained optimization}\label{constrained}

Again, let $f$ be a $d$-dimensional load distribution and say we want to again
partition it into $k = (k_1, \dots, k_d)$ parts. This time however, we will have
equality constraints among the partitions, e.g., $p_1 = p_2 = p_3$ and $p_4 =
    p_d$, etc. Let's say we group the dimensions whose partitions are constrained to
be equal and we are left with only $\hat d$ groups. Then, we can encode these
constraints as:
\begin{align*}
    p_i        & = \hat p_{\hat i},\text{ for some } \hat i \in [\hat d]^+ \\
    C_{\hat i} & = \{i \mid p_i = \hat p_{\hat i}\}
\end{align*}

If we optimize over $\hat p$, then the problem turns into an unconstrained
one. However, first we need to find a way to parametrize $\hat p$ as $\hat
    \pi$ in a similar manner to our former discussion. First, we
define $\hat F_{\hat i}(x)$ as:
$$ 
    \hat F_{\hat i}(x_{\hat i}) = \frac{1}{|C_{\hat i}|}\sum_{i \in C_{\hat i}} F_i(x_i)
$$ 
With this, we parametrize $\hat p$ as $\hat \pi$ as in the previous section:
$$ 
    \hat p_{\hat i}[j_{\hat i}] = \hat F^{-1}_{\hat i}(\hat \pi_{\hat i}[j_{\hat i}])
$$ 
As is the case with $p_i$ and $\hat p_{\hat i}$, we will define $\pi_i$ as an
alias to the corresponding $\hat \pi_{\hat i}$. In a similar manner, we can
define $\hat r_{\hat i}(f, \hat \pi)[j_{\hat i}]$ as:
\begin{equation}
    \hat r_{\hat i}(f, \hat \pi)[j_{\hat i}] = \max_{\forall i \in C_{\hat i}} r_i(f, \pi)[j_{\hat i}]
    \label{hat_ri_computation}
\end{equation}
and finally, we can define the subgradient, again, the same as in the previous section:
\begin{equation}
    \hat g_{\hat i}[j_{\hat i}] = \sum_{u = 0}^{j_{\hat i} - 1} \hat r_{\hat i}[u] - \frac{j_{\hat i}}{k_{\hat i}} \sum_{u = 0}^{k_{\hat i} - 1} \hat r_{\hat i}[u], \forall i, j_{\hat i}
    \label{subgradient_computation}
\end{equation}

\subsection{Initializing \texorpdfstring{$\hat \pi$}{{pi}}}
\label{initialization}

The possible values for $\hat \pi_{\hat i}[j_{\hat i}]$ lie in the range of $\hat F_{\hat i}$.
Since $\hat F_{\hat i}$ is a monotonic function and $\hat F_{\hat i}(-\infty) =
    0$ and $\hat F_{\hat i}(\infty) = L(f)$ for all $\hat i$, we could choose to
deterministically initialize $\hat \pi_{\hat i}[j_{\hat i}]$ as:
$$ 
    \hat \pi_{\hat i}[j_{\hat i}] = \frac{j_{\hat i}}{\hat k_{\hat i}} L(f).
$$ 

Another option to initialize $\hat \pi_{\hat i}[j_{\hat i}]$ is to use the uniform distribution
with range $(0, L(f))$. Note that $\hat \pi_{\hat i}$ has to be monotonic so we sort
$\hat \pi_{\hat i}$ after initializing them with the uniform distribution. After this,
we set $\hat \pi_{\hat i}[0] = 0$ and $\hat \pi_{\hat i}[\hat k_{\hat i}] = L(f)$. This way of
initialization is better because we suspect that $d$-dimensional rectilinear
partitioning for any $d \geq 2$ and any types of constraints is NP-hard. Thus,
it is expected that there are many local optima. Random initialization makes it
so that multiple runs of the algorithm with different random seeds produce
different outputs which can be considered to be a good property since a single
run might get stuck at a bad local optima.

\subsection{Stopping condition}
\label{stopping}

Since we have characterized the optimal solution to be in the set $S$ defined in~\eqref{optimality}, this immediately gives us a metric to decide when to
stop. Since the following holds
$$ 
    L_\pi(f, \pi) = \max_{j_{\hat i}} \hat r_{\hat i}(f, \hat \pi)[j_{\hat i}], \forall \hat i
$$ 
we can measure for all $\hat i$ how close $r_{\hat i}(f, \hat \pi)$ is to the uniform
distribution when we consider $r_{\hat i}(f, \hat \pi)$ as an unnormalized probability distribution. This can be
done using norms, including $L_1$, $L_2$ or even $L_\infty$. During
the iterations, we can check whether it is close enough and if so, we can choose
to terminate the algorithm. Among all of the available options, we choose to stop the
algorithm when the following holds: \begin{align}
    \frac{L_\pi(f, \pi) - \min_{\hat i, j_{\hat i}} \hat r_{\hat i}[j_{\hat i}]}{\min_{\hat i, j_{\hat i}} \hat r_{\hat i}[j_{\hat i}]} < \epsilon
    \label{local_optimality_measure}
\end{align}
However, when we represent sparse tensors as load
distributions, the load function that we defined in
\eqref{load_definition} won't be continuous. Thus, it might be impossible for
$\hat r_{\hat i}(f, \hat \pi)$ to approximate an unnormalized uniform probability distribution causing
the algorithm to never stop. Therefore, we resort to the following technique:
for $c \sum_{i=1}^d k_i$ iterations after the last update of the best
solution found so far, if the solution quality doesn't improve more than a factor of
$1 + \epsilon$, then \sgo stops. Our experimental results show that combining
this technique with the aforementioned stopping condition gives good results with
$c = 10$ and $\epsilon = 0.001$.

\begin{figure*}[htb]
    \begin{subfigure}[b]{0.45\textwidth}
        \centering
        \includegraphics[width=\textwidth]{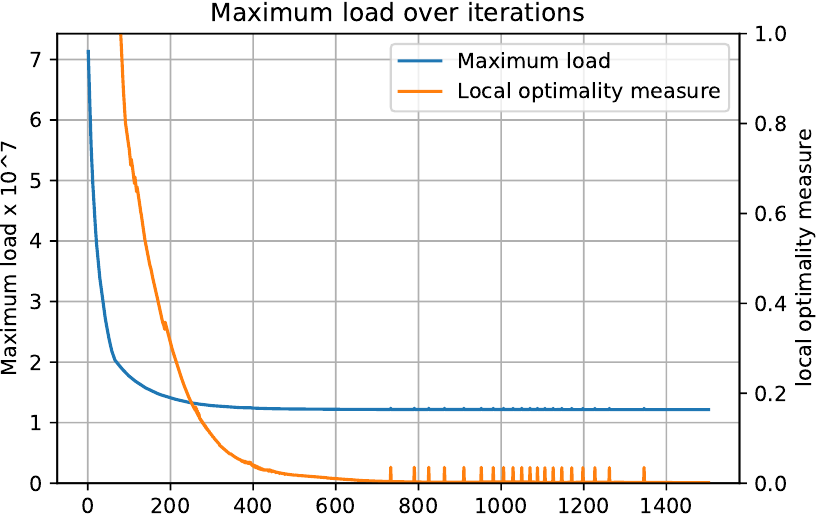}
        \caption{{\small $(16, 16)$ rectilinear partitioning}}\label{fig:twitter7_rect}
    \end{subfigure}
    \hfill
    \begin{subfigure}[b]{0.45\textwidth}
        \centering
        \includegraphics[width=\textwidth]{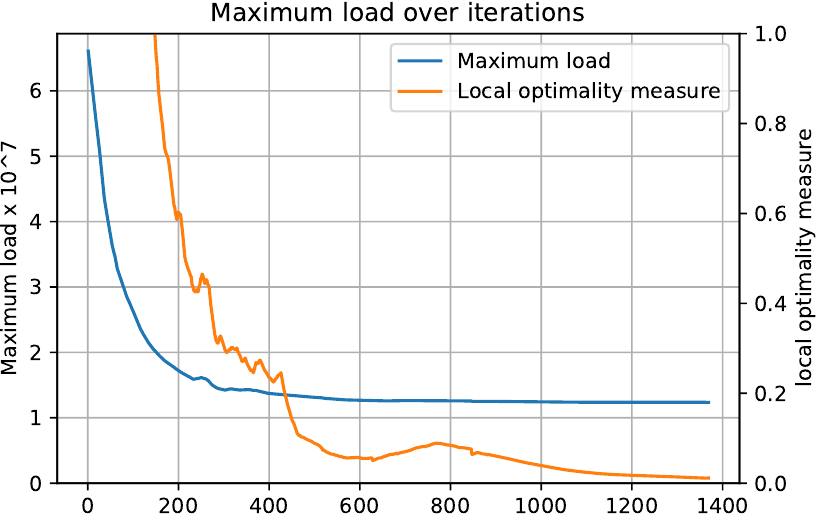}
        \caption{{\small $(16, 16)$ symmetric rectilinear partitioning}}\label{fig:twitter7_sym}
    \end{subfigure}
    \caption{\small Plots of $L(f, \pi)$ and closeness to local optimality with respect to iteration number on twitter7 matrix from SuiteSparse with $(16, 16)$ partitioning.}
    \label{fig:twitter7}
\end{figure*}

In Figure~\ref{fig:twitter7}, if the orange curve gets sufficiently close to
$0$, then we can stop because we have reached a local optima. We can also choose
to stop when the blue curve starts to flatten in case we can't get close enough
to a local optima. In Figure~\ref{fig:twitter7_sym}, we observe that both the blue
and orange curves are less smooth compared to Figure~\ref{fig:twitter7_rect}.
Our explanation for this phenomenon is that
the parameters of partition arrays of both dimensions are shared in the
symmetric case, so the effect of the first and the second dimensions to the subgradient
sometimes conflict each other. At those times, \sgo might move $\hat \pi$ further
away from the set $S$ defined in~\eqref{optimality}. Another reason for non-smoothness is
that the object we are partitioning is discrete in nature. That is why there are some
jumps at the end of the blue and orange curves in Figure~\ref{fig:twitter7_rect}.

\subsection{Step size selection}

The update rule we defined in \eqref{update_rule} depends on current
iteration $t$. There is a multitude of step size rules that can be used one of
which is the constant step size rule~\cite{boyd2003subgradient}. However, we choose to use a diminishing step
size rule, specifically $\eta_i(t) \approx \frac{\mu}{\sqrt{\frac{t}{\hat k_i} +
            T}}$, where $\mu = 1$ and $T = 100$ were determined to work well empirically
in our experiments.

\subsection{Overall summary and an example}

\begin{figure*}
    \centering
    \subcaptionbox{{\small Initial partitioned matrix}\label{fig:part_example_mat}}{\includegraphics[width=0.31\textwidth]{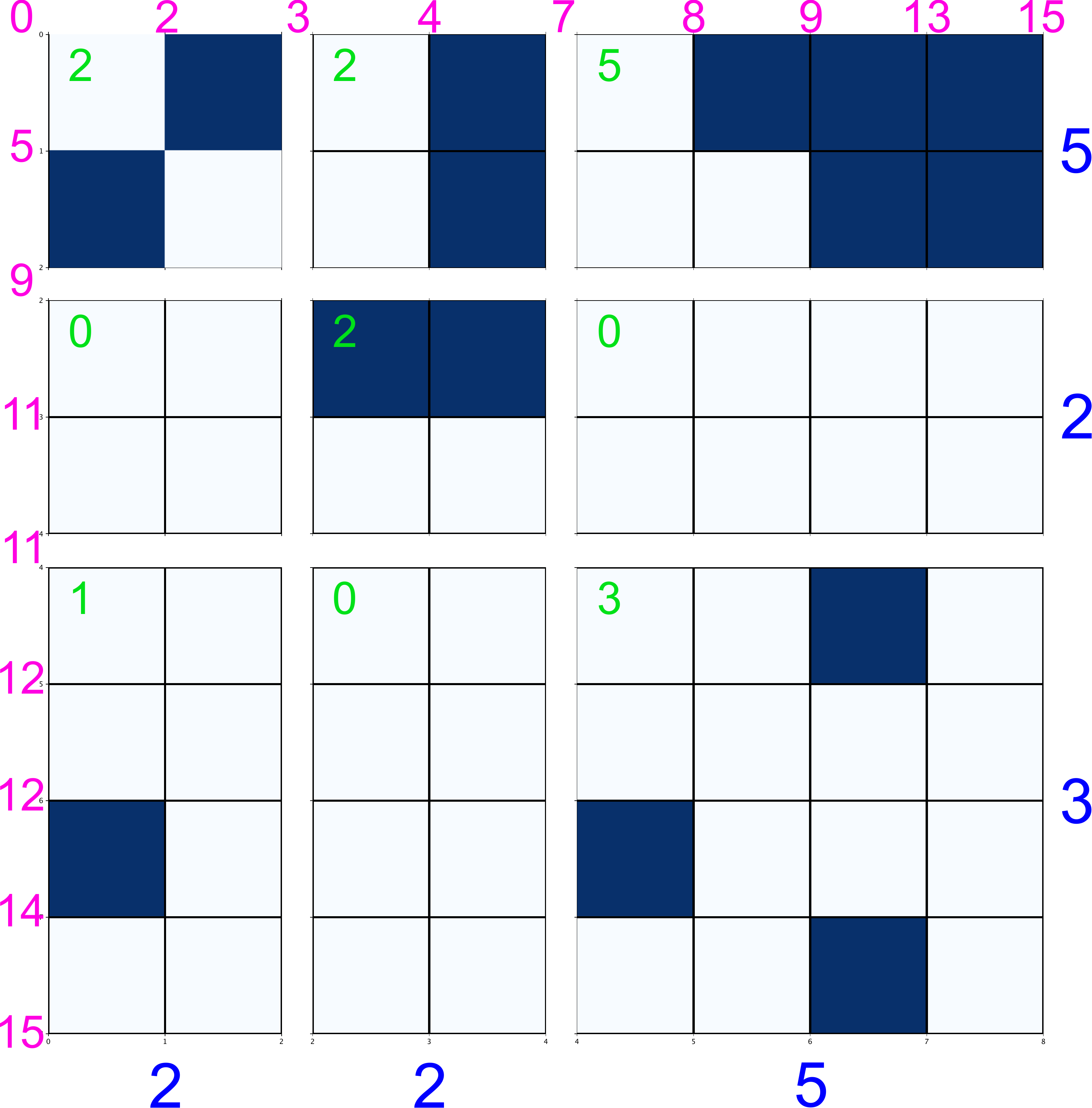}}
    \hfill
    \subcaptionbox{{\small After a single update step of RPP}\label{fig:part_mat_example_RPP}}{\includegraphics[width=0.31\textwidth]{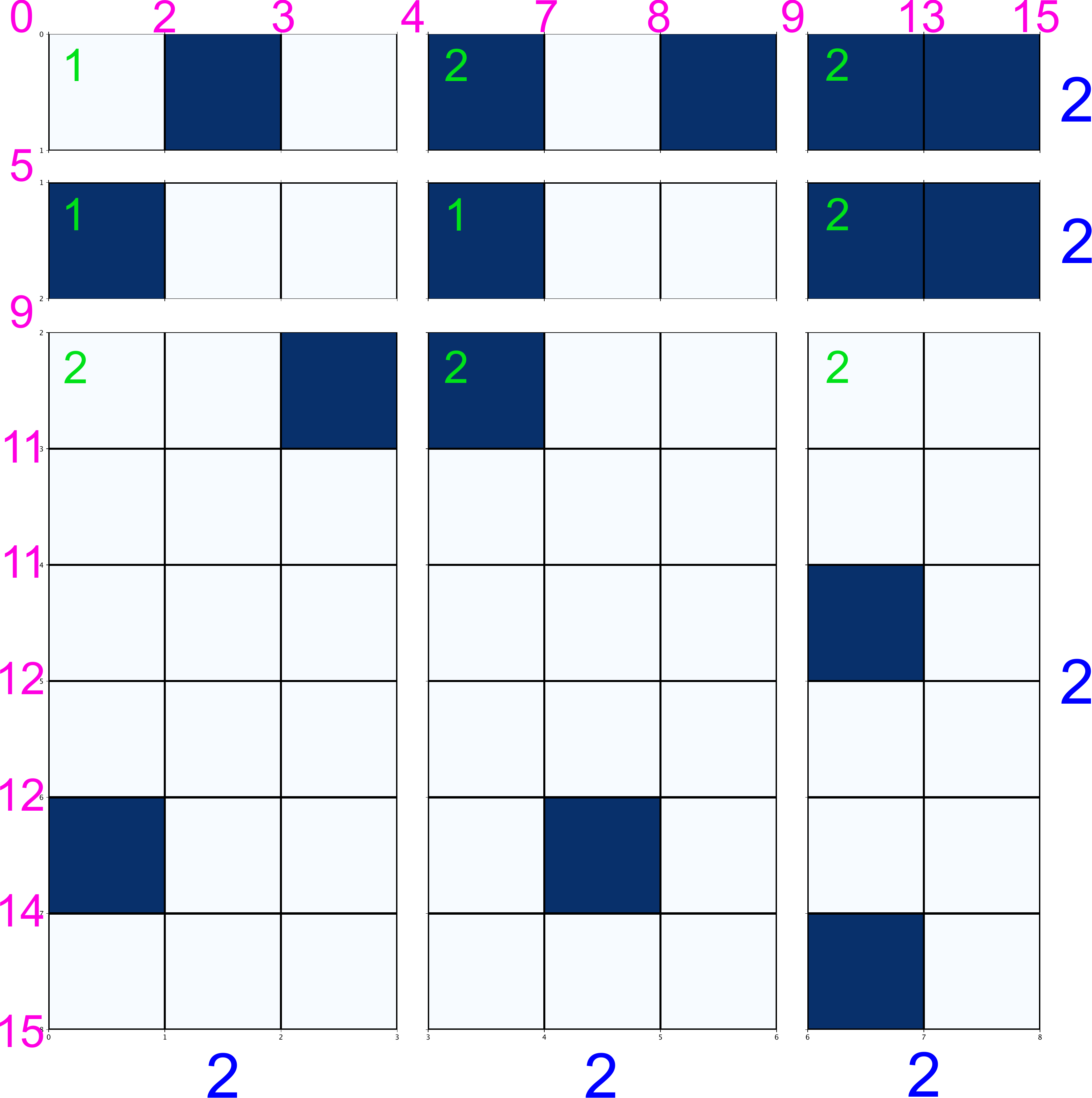}}
    \hfill
    \subcaptionbox{{\small After a single update step of SRPP}\label{fig:part_example_mat_SRPP}}{\includegraphics[width=0.31\textwidth]{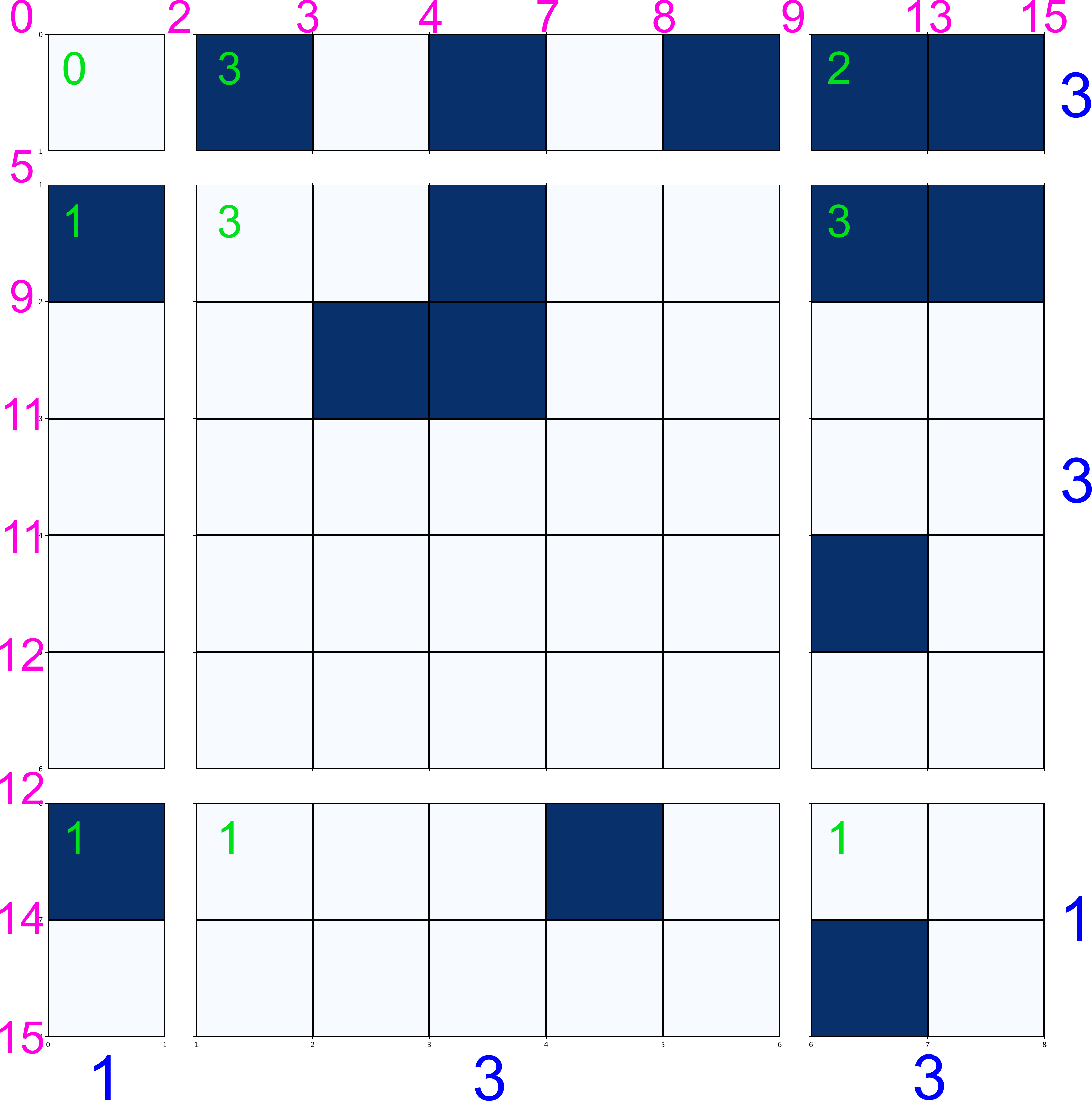}}
    \caption{\small A toy matrix partitioned in various ways. The (magenta) numbers to the left and
        top of the figures denote the prefix sums $F_1$ and $F_2$. The (blue) numbers to the right
        and bottom represent the maximum loads $r_1$ and $r_2$.
        Finally, the (green) numbers inside the boxes show the load of the part they are in.}
    \label{fig:part_example}
\end{figure*}

Algorithm~\ref{alg:sgo} presents the pseudocode of our proposed method \sgo.
First, we initialize the partitioning variables $\hat \pi$. Then, in each step,
\sgo computes the subgradients $\hat g_{\hat i}$ and updates $\hat \pi$, and
keeps track of the best solution found so far, $\hat \pi^*$. \sgo returns the
best solution when the stopping condition is achieved.

\begin{algorithm2e}[ht]
    \tcp*[h]{$f$: A $d$-dimensional load distribution}\;
    \tcp*[h]{$C$: Set of constraints}\;
    \tcp*[h]{$k = (k_1,\dots, k_d)$: Partition vector sizes of each dimension}\;

    $\hat \pi$ = {\sc Initialize}()\tcp*[r]{See Section~\ref{initialization}}
    $\hat \pi^* = \hat \pi$ \tcp*[r]{Initialize best solution found so far}
    $t = 0$ \tcp*[r]{Initialize iteration count to $0$}
    \kc{See Section~\ref{stopping} for the stopping condition}
    \While{{\bf not} \eqref{local_optimality_measure} }{
        Compute $r_i$ using~\eqref{ri_computation}, $\hat r_{\hat i}$ using~\eqref{hat_ri_computation},
        and $\hat g_{\hat i}$ using~\eqref{subgradient_computation}\;
        $\hat \pi_{\hat i} = \hat \pi_{\hat i} - \eta_{\hat i}(t) \hat g_{\hat i}$ st. ${\hat i}\in [d]^+$\;
        \If{$L(f, \hat \pi) < L(f, \hat \pi^*)$}{
            $\hat \pi^* = \hat \pi$\tcp*[r]{Best solution improved}
        }
        $t = t+1$\;
    }

    \Return $\hat \pi^*$

    \caption{{\sc \sgo}($f$, $C$, $k$)}
    \label{alg:sgo}
\end{algorithm2e}

As a toy example,
given the partitioned matrix $A$ in Figure~\ref{fig:part_example_mat}
as our initial state with partitioning $(p, p)$,
where $p = [0, 2, 4, 8]$, we apply a single update of our algorithm when
there are no constraints. Since $A$ is a sparse matrix, we first get its load
distribution $f_A$. Note that $L(f_A, (p, p)) = 5$. After that, we compute the
prefix sums $F_1$ and $F_2$ by counting the numbers of nonzeros along the rows
and columns to get $F_1 = [0, 5, 9, 11, 11, 12, 12, 14, 15]$ and $F_2 = [0, 2,
    3, 4, 7, 8, 9, 13, 15]$. By plugging $p$ into $F_1$ and $F_2$ as an index,
we get $\pi_1 = [F_1[p[0]], F_1[p[1]], F_1[p[2]], F_1[p[3]]] = [0, 9, 11, 15]$
and $\pi_2 = [F_2[p[0]], F_2[p[1]], F_2[p[2]], F_2[p[3]]] = [0, 3, 7, 15]$. We also compute the
loads of each part to get $[[2, 2, 5], [0, 2, 0], [1, 0, 3]]$. As the next step,
we compute $r_1 = [5, 2, 3]$ and $r_2 = [2, 2, 5]$. After that, we compute the
subgradients $g_1 = [0, \frac{5}{3}, \frac{1}{3}, 0]$ and $g_2 = [0, 1, -1, 0]$.
If we have the step size $\eta = 2$, then updated parameters become $\pi_1 = [0,
    \frac{17}{3}, \frac{31}{3}, 15]$ and $\pi_2 = [0, 5, 11, 15]$. Doing binary
searches in $F_1$ and $F_2$ to compute $F_1^{-1}$ and $F_2^{-1}$, we get the new
partition vectors $p_1 = [0, 1, 2, 8]$ and $p_2 = [0, 3, 6, 8]$. With this new
partition of the matrix, we have $L(f_A, (p_1, p_2)) = 2$, down from $5$.

Now we show a single step of our algorithm when there is an equality
constraint along the matrix rows and columns. In this case, we compute $\hat F_1
    = \frac{F_1 + F_2}{2} = [0, \frac{7}{2}, 6, \frac{15}{2}, 9, 10, \frac{21}{2},
    \frac{27}{2}, 15]$. By looking-up each element of $p$ in $\hat F_1$, we
get $\hat \pi_1 = [0, 6, 9, 15]$. Since we already computed $r_1$ and $r_2$
above, we can compute $\hat r_1 = [5, 2, 5]$ by taking an element-wise max of
$r_1$ and $r_2$. After that, we compute the subgradient $\hat g_1 = [0, 1, -1,
    0]$. Using $\eta = 2$, we get updated $\hat \pi = [0, 4, 11, 15]$.
Doing binary searches on $\hat F_1$, we get the new partition vector $\hat p_1 =
    [0, 1, 6, 8]$. Because of the constraint $p_1 = \hat p_1 = p_2$, we have
$L(f_A, (p_1, p_2)) = 3$, down from $5$.

\section{Mapping Partitioning Problems into \sgo}
\label{sec:mapping}

In this section, we propose modeling strategies to map four different applications
into \sgo. The first two of these applications model the partitioning problem using
two-dimensional objective functions, while the last two of these applications use
three-dimensional objective functions.

\subsection{2-dimensional RPP}
\label{rectilinear}

Nicol's~\cite{Nicol94-JPDC} rectilinear partitioning algorithm partitions a
given sparse matrix $A$ into $(k_1, k_2)$ parts and tries to minimize the maximum
number of nonzeros contained in the most loaded partition. \sgo can achieve the same type of
partitioning: Given a $2$-dimensional sparse matrix
$A$, we treat it as a $2$-dimensional load distribution $f_A$ as in~\eqref{sparse_tensor}.
The objective is to partition $f_A$ into $(k_1, k_2)$
parts. Thus, this use of our framework is a direct contender to Nicol's
algorithm, which we will investigate in the experiments section. We will refer
to this variant of \sgo as \sgotwdr. In short, we will have \sgo solve the following
optimization problem:
\begin{equation}
    \min_{p_1, p_2} \max_{j_1, j_2} L(f_A, (p_1, p_2), (j_1, j_2))
    \label{obj_rect}
\end{equation}

\subsection{2-dimensional SRPP}
\label{symmetric}

The partitioning algorithms presented in~\cite{Yasar21-JEA} partition a given
sparse matrix $A$ into $(k, k)$ parts resulting in partition vectors $(p, p)$ while
minimizing the maximum number of nonzeros contained in a single partition. Note that
the use of $p$ for the partition vectors of both dimensions implies an equality
constraint, as explained in Section~\ref{constrained}. The objective is to
partition $f_A$ into $(k, k)$ parts. This kind of use of our framework is a
direct contender to the algorithms presented in~\cite{Yasar21-JEA}.
In our experiments, we compare \sgo with the PAL algorithm implemented in that library.
We will refer to this variant of \sgo as \sgotwds. In short, \sgo solves
the following optimization problem:
\begin{equation}
    \min_{p_1 = p_2} \max_{j_1, j_2} L(f_A, (p_1, p_2), (j_1, j_2))
    \label{obj_sym}
\end{equation}

\subsection{3-dimensional RPP Use Case: SpGEMM}
\label{nsym_spgemm}

\sgo is a flexible framework, and it can optimize different objective functions.
This property is highly useful for modeling a wide range of applications.
For instance, for the Sparse Matrix-Matrix Multiplication (SpGEMM) kernel that
uses the SUMMA algorithm, one might want to
minimize the maximum communication volume during each communication round. In this
algorithm, each processor $(u, v)$ in a $k \times k$ processor grid multiplies the
tile $(u, w)$ of $A$ with the tile $(w, v)$ of $B$ and adds it to the tile $(u, v)$
of C in communication round $w$. The total volume of communication done by the processor
$(u, v)$ in round $w$ is the sum of the load of the tile $(u, w)$ of $A$ and the tile
$(w, v)$ of $B$. The goal is to minimize the maximum total
volume of communication in the round $w$ between all processors, i.e. $\max_{u, v}
    nnz(A[u, w]) + nnz(B[w, v])$. When we consider all of the communication rounds,
the objective becomes to minimize $\sum_w \max_{u, v} nnz(A[u, w]) + nnz(B[w, v])$.
However, it is not possible to optimize this objective function with our framework as
subgradients vanish when we sum over a dimension. That is why we choose to minimize
$\max_{u, w, v} nnz(A[u, w]) + nnz(B[w, v])$.

To map this problem into our framework, let $f_A$ and $f_B$ be
$2$-dimensional load distributions representing matrices $A$ and $B$ as in
Section \ref{spmtx}. Let $f(x_1, x_2, x_3) = f_A(x_1, x_2) + f_B(x_2, x_3)$.
Note that $f$ is $3$-dimensional load distribution. Solving the problem of
partitioning $f$ into $(k, k, k)$ parts and getting the resulting partitions $(p_1,
    p_2, p_3)$, directly corresponds to minimize the communication
volume of the Sparse SUMMA algorithm where $A$ is distributed with respect to
$(p_1, p_2)$ and $B$ is distributed with respect to $(p_2, p_3)$. In short,
\sgo solves the following optimization problem:
\begin{equation}
    \min_{p_1, p_2, p_3} \max_{j_1, j_2, j_3} L(f, (p_1, p_2, p_3), (j_1, j_2, j_3))
    \label{obj_spgemm}
\end{equation}

We will refer to this variant of \sgo as \sgothdr.

\subsection{\texorpdfstring{$3$}{3}-dimensional SRPP Use Case: Triangle Count}
\label{sym_tri}

Triangle Counting Problem~\cite{Hu18-SC,Yasar22-TPDS} can be another use-case.
When we partition the adjacency matrix of a given graph using a symmetric-rectilinear
fashion, edges of a triangle can appear in at most three of the partitions.
Furthermore, in a triangle, for chosen two edges, there exists
only one sub-graph such that the third edge belongs.
If one partitions the adjacency matrix $A$
into $(k, k)$ parts using a partition $p_1, p_2$ under the constraint $p_1 =
    p_2$, then there will be $\approx k^3$ instances (i.e., tasks) of the triangle counting problem, each
seeks triangles in $(A[u, w], A[w, v], A[u,v])$. In the heterogeneous environment,
where the memory of the co-processors or the bandwidth between the host-processor
and the co-processors are limited, $nnz(A[u, w]) + nnz(A[w, v]) + nnz(A[u, v])$ of
task $(u, w, v)$ will correspond to transfer times to the co-processors and the memory
requirement. Thus, the objective is to minimize
$\max_{u, w, v} nnz(A[u, w]) + nnz(A[w, v]) + nnz(A[u, v])$.

To map this problem into our framework, let $f_A$ be a $2$-dimensional
load distribution representing the adjacency matrix $A$ and let $f(x_1, x_2,
    x_3) = f_A(x_1, x_2) + f_A(x_2, x_3) + f_A(x_1, x_3)$. Solving the problem of
partitioning $f$ using $(p_1, p_2, p_3)$ into $(k, k, k)$ parts with the
constraint $p_1 = p_2 = p_3$, directly corresponds to minimize
the maximum communication cost of a processor. We will
refer to this variant of \sgo as \sgothds. In short, \sgo solves the following
optimization problem:
\begin{equation}
    \min_{p_1 = p_2 = p_3} \max_{j_1, j_2, j_3} L(f, (p_1, p_2, p_3), (j_1, j_2, j_3))
    \label{obj_tri}
\end{equation}

\section{Experiments}
\label{sec:experiments}

In this section, we compare the performance of our proposed algorithms
(\sgotwds, \sgotwdr, \sgothdr, \sgothds) with
state-of-the-art rectilinear and symmetric rectilinear partitioning algorithms.
We use Nicol's (NIC)~\cite{Nicol94-JPDC}, Aspvall et al.'s (2SWP)~\cite{Aspvall01-TCS} and Muthukrishnan and Suel's (4APX)~\cite{muthu2005} rectilinear partitioning
algorithms for
the nonsymmetric case and Probe a Load (PAL) symmetric rectilinear partitioning
algorithm for the symmetric case. We use
NIC and PAL with their default parameters from the SARMA library~\cite{Yasar21-JEA}.
We choose to use a maximum iteration limit of 10000 for the 4APX algorithm with $\epsilon = 0.01$.
For the 4APX algorithm, the authors state in their paper that the number of iterations
required to converge is on the order of
$O(k_1 \log N)$, where $k_1$ stands for the number of parts in a dimension and $N$ stands
for the maximum dimension of the input matrix.
We also include uniform partitioning (UNI) in our experiments as the baseline.
Note that \sgotwds and \sgothds output symmetric partitions whereas \sgotwdr and \sgothdr
output rectilinear partitions.

\sgo variants use the random initialization as explained in Section~\ref{initialization}.
To reduce the variance caused by randomness, the median result of 10 runs is taken in all
reported results.

We ran all of the experiments on the Hive cluster of Georgia Tech. Hive has
$416$ compute nodes, each is equipped with $2 \times 2.7$ GHz Intel Xeon
$6226$ CPUs (with 12-cores), and $192$ GB of RAM. Interconnection network is
EDR Infiniband (100Gbps). Each algorithm run had a single such node with all 24
cores for their use.

All of the sparse matrices used in our
experiments were downloaded from the SuiteSparse Matrix Collection
\cite{Davis11-TOMS}. We excluded non-square matrices and matrices with less than
$10^6$ or more than $2 \times 10^9$ nonzeros. By the time of this experimentation
there were $687$ matrices, out of $2856$, that fit our criteria.

We downloaded $17$ additional point datasets from the DIMACS10 workshop repository~\cite{DIMACS10};
Street Networks, and Frames from 2D Dynamic Simulations categories. The number of points
in this dataset varies from $10^5$ to $5 \times 10^7$.

We present our results using performance profile plots~\cite{Dolan02-MP}. In
these plots, the $y$-axis denotes the relative number of test instances,
and the $x$-axis denotes the ratio of the metric of interest to the best
performing algorithm on one of the test instances.
The higher and closer a plot is to the $y$-axis, the better the method is.

Furthermore, in order to support reproducibility, we provide normalized load imbalance
(with respect to average non-zero per part) and absolute algorithm execution times
for a subset of data in Appendix~\ref{sec:appendix}.

\subsection{Implementation}

We have contributed our implementation of \sgo, the two-sweep (2SWP) algorithm in~\cite{Aspvall01-TCS}
and the four approximation (4APX) algorithm in~\cite{muthu2005} to the SARMA library
and it is publicly available at \url{https://github.com/GT-TDAlab/SARMA} via a BSD-license.
SARMA library is a suite of
spatial partitioning algorithms implemented using C++17 and the parallel standard library
using shared memory parallelism.
We particularly used the sparse prefix sum data structure provided in SARMA to
represent $2$-dimensional load distributions
$f_A$ implied by the sparse matrices $A$ used during experiments. Given a sparse
matrix $A$ of dimensions $(m, n)$ with $o$ nonzeros, this data structure enables
us to query the load of a rectangular region in the $2$-dimensional space in
$O(\log n \log m)$ time using $O(o \log \min(m, n))$ space.

The computational complexity of our algorithm is given by sparse-prefix-sum data
structure construction and load queries. \sgo can
be computed in $O(o \log \min(m, n) + \tau k_1 k_2 \log n
    \log m)$ for the $2$-dimensional case where $\tau$ stands for the number of
iterations. Note that, while the number of iterations required depends on many
factors, it is at most thousands in practice and it highly depends on
the selected parameters for the stopping condition and the step size.
The data structure construction dominates the complexity of our algorithm. However, one can reduce
that by sparsifying the graph, that is, sampling the nonzeros of the input sparse-matrix~\cite{Yasar21-JEA}.

\subsection{Evaluation of the Partitioning Quality}
\label{subsec:experimental-load}

In this sub-section, we compare the performance of our proposed algorithms
with-respect-to state-of-the-art partitioning algorithms for given objective
functions. Depending on the nature of the partitioning problem we use
NIC, 4APX, 2SWP, PAL, and UNI algorithms as baselines. In the following experiments
we present results for $8\times 8$, $16\times, 16$, and $32\times 32$
partitionings.

\subsubsection{\texorpdfstring{$2$}{2}-dimensional Rectilinear Partitioning}
\label{subsubsec:twdr}

\begin{figure*}[htb]
    \begin{subfigure}[b]{0.3\textwidth}
        \centering
        \includegraphics[width=\textwidth]{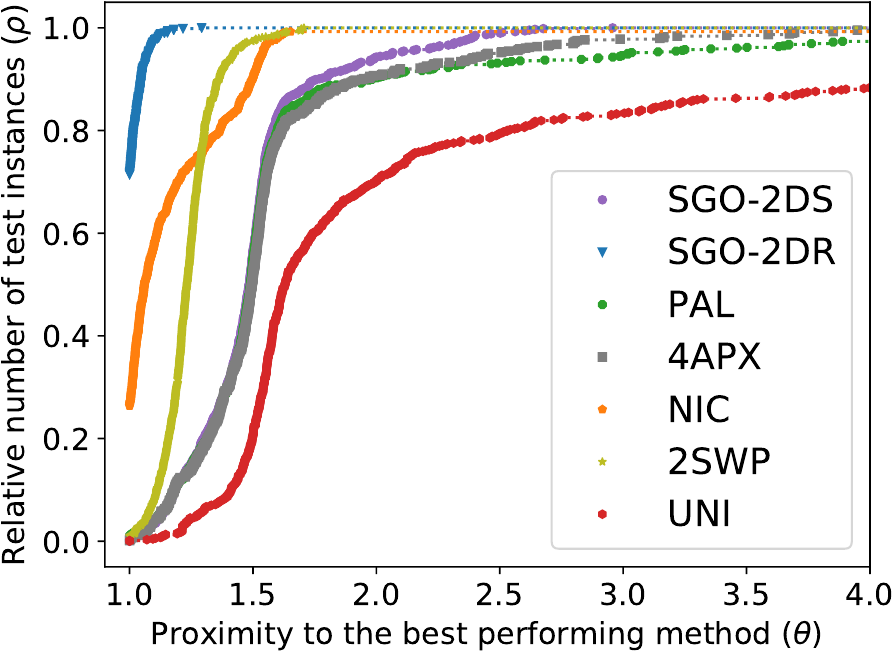}
        \caption{{\small $(8, 8)$}}
        \label{fig:rect_perf_8}
    \end{subfigure}
    \hfill
    \begin{subfigure}[b]{0.3\textwidth}
        \centering
        \includegraphics[width=\textwidth]{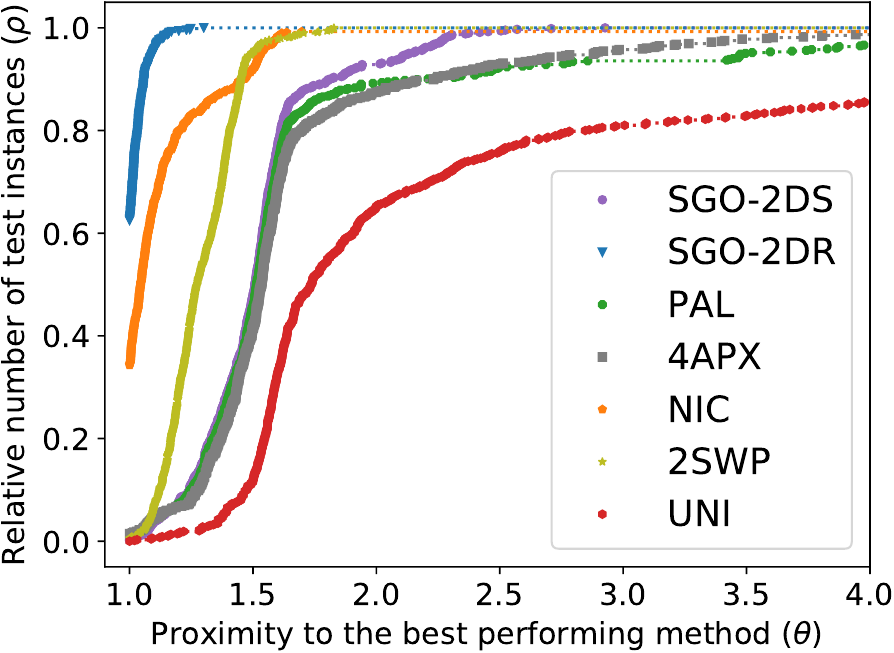}
        \caption{{\small $(16, 16)$}}
        \label{fig:rect_perf_16}
    \end{subfigure}
    \hfill
    \begin{subfigure}[b]{0.3\textwidth}
        \centering
        \includegraphics[width=\textwidth]{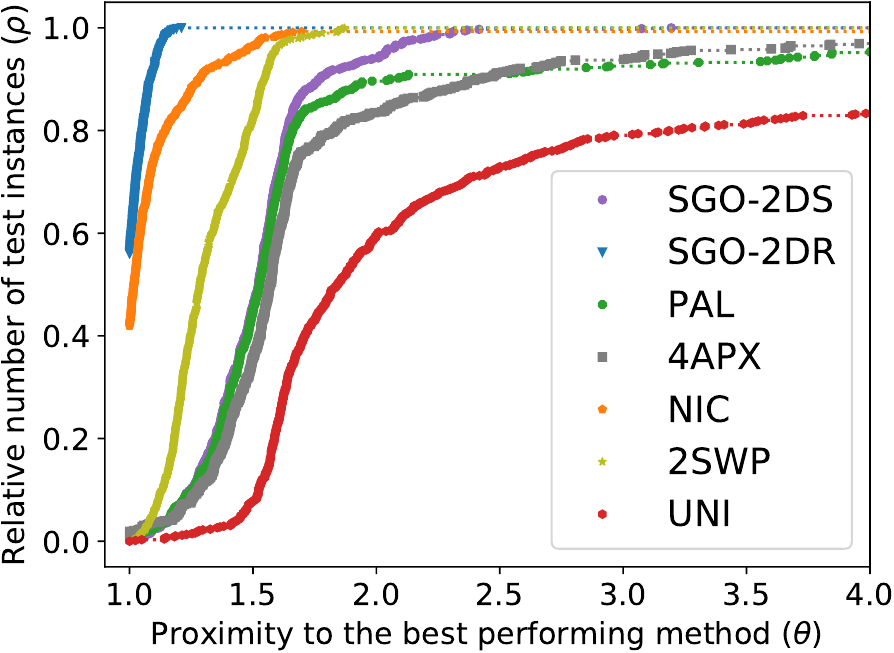}
        \caption{{\small $(32, 32)$}}
        \label{fig:rect_perf_32}
    \end{subfigure}
    \caption{\small Performance profile plots of the partitioning methods
        with natural reordering. The algorithms are compared wrt.~\eqref{obj_rect}.}
    \label{fig:rect_perf}
\end{figure*}

In this experiment, we compare \sgotwdr, and \sgotwds with NIC, 4APX, 2SWP,
PAL and UNI
algorithms. In this experiment the objective function is minimizing the load of
the maximum loaded partition. Note that comparing SRPP algorithms with RPP
algorithms is not fair because SRPP algorithms have more constraints. However we
include these algorithms in this experiment to give the reader an idea of how much
of a limitation SRPP brings compared to RPP. Figure~\ref{fig:rect_perf} illustrates
that the relative order of the algorithms with respect to partitioning quality
is \sgotwdr, NIC, 2SWP, \sgotwds, PAL, 4APX and UNI. We see that the difference between
\sgotwdr and NIC algorithms start to decrease as we increase the number of parts
from $(8, 8)$ to $(32, 32)$. However, \sgo still outperforms NIC both in terms of
partitioning quality and also execution time as we will present below.
Since RPP algorithms can output different partition
vectors for each dimension as opposed to SRPP algorithms, this kind of a difference
was expected.

\begin{figure*}[htb]
    \begin{subfigure}[b]{0.3\textwidth}
        \centering
        \includegraphics[width=\textwidth]{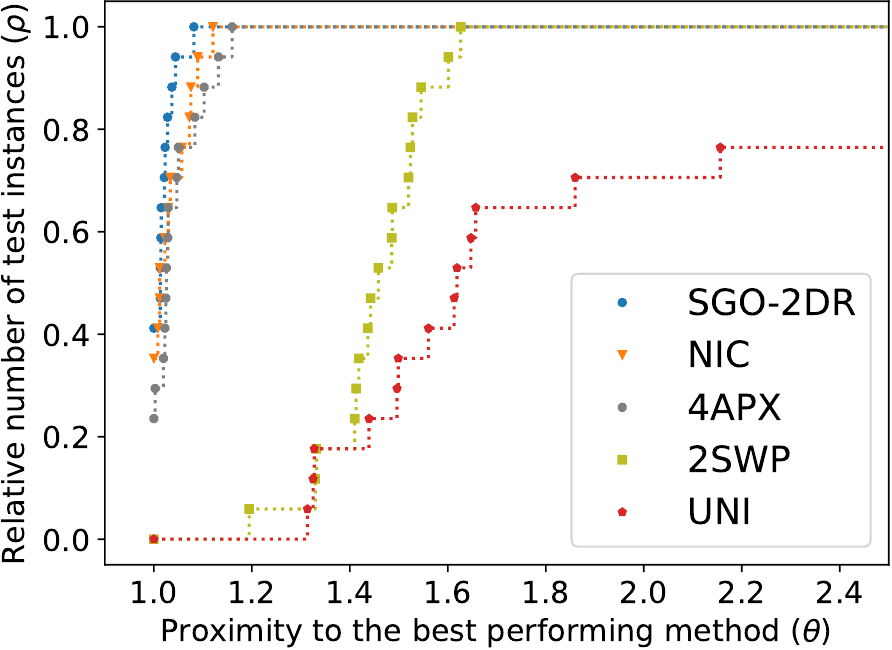}
        \caption{{\small $(8, 8)$}}
        \label{fig:spat_perf_8}
    \end{subfigure}
    \hfill
    \begin{subfigure}[b]{0.3\textwidth}
        \centering
        \includegraphics[width=\textwidth]{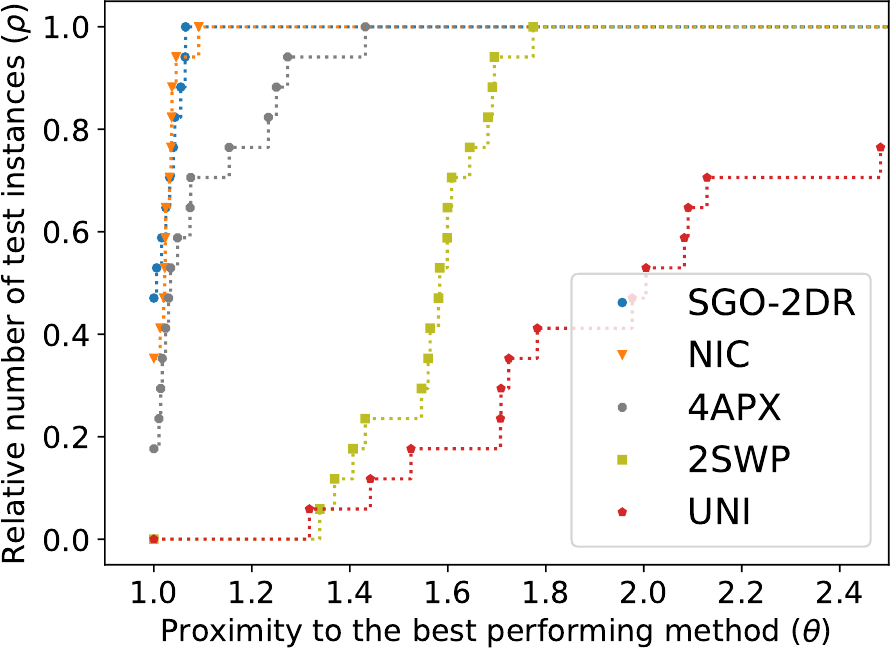}
        \caption{{\small $(16, 16)$}}
        \label{fig:spat_perf_16}
    \end{subfigure}
    \hfill
    \begin{subfigure}[b]{0.3\textwidth}
        \centering
        \includegraphics[width=\textwidth]{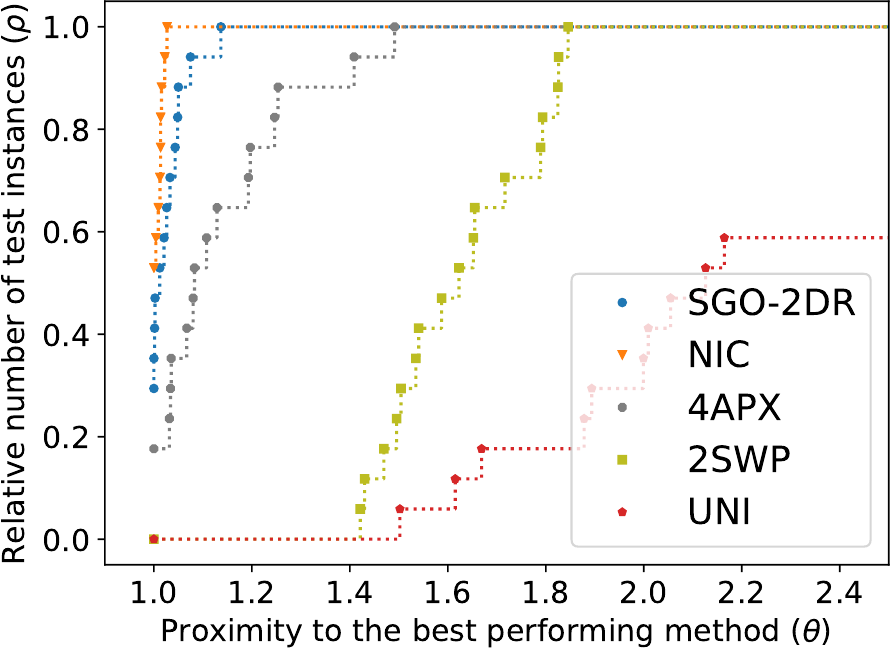}
        \caption{{\small $(32, 32)$}}
        \label{fig:spat_perf_32}
    \end{subfigure}
    \caption{\small Performance profile plots of the partitioning methods
        on point datasets. The algorithms are compared wrt.~\eqref{obj_rect}.}
    \label{fig:spat_perf}
\end{figure*}

In the point datasets, we see that \sgotwdr and NIC are much closer, \sgotwdr outperforming NIC for the $(8, 8)$ case and NIC
outperforming \sgotwdr for the $(32, 32)$ case in Figure~\ref{fig:spat_perf}. Note that, the number of instances
is small, only 17. Thus, we believe that the use of sparse matrices gives a better picture
of overall quality.

\begin{figure*}[htb]
    \begin{subfigure}[b]{0.3\textwidth}
        \centering
        \includegraphics[width=\textwidth]{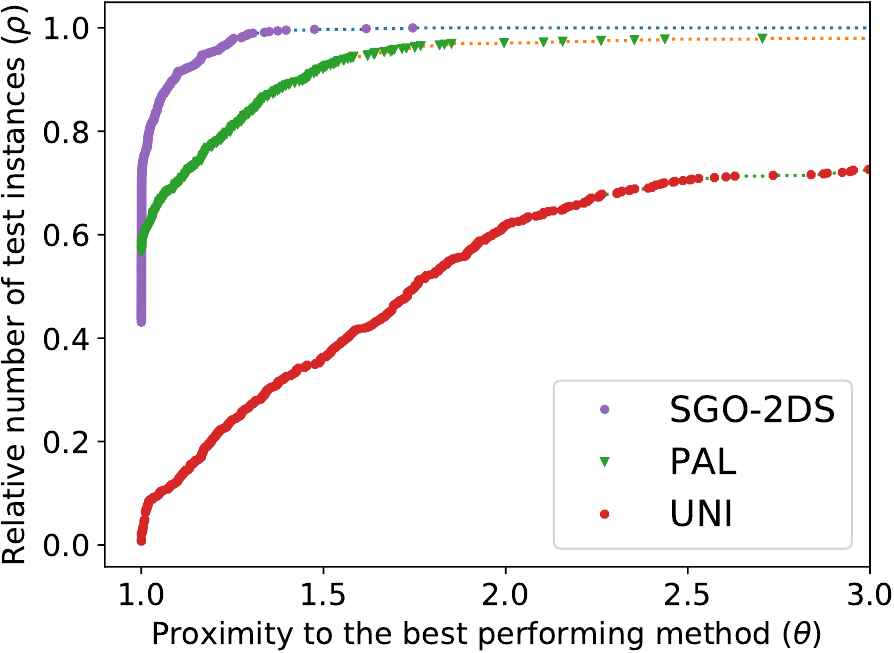}
        \caption{{\small $(8, 8)$}}
        \label{fig:asc_sym_perf_8}
    \end{subfigure}
    \hfill
    \begin{subfigure}[b]{0.3\textwidth}
        \centering
        \includegraphics[width=\textwidth]{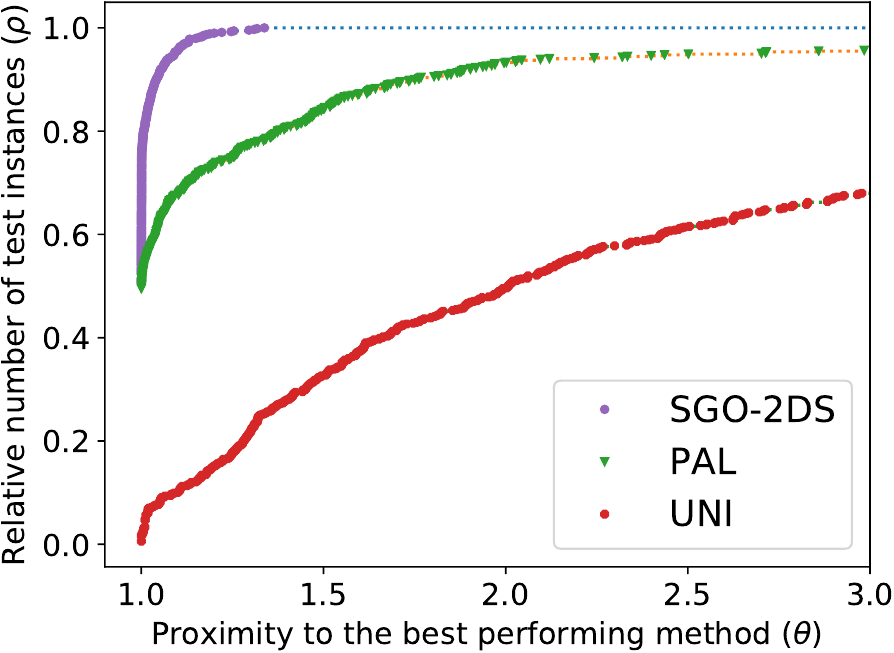}
        \caption{{\small $(16, 16)$}}
        \label{fig:asc_sym_perf_16}
    \end{subfigure}
    \hfill
    \begin{subfigure}[b]{0.3\textwidth}
        \centering
        \includegraphics[width=\textwidth]{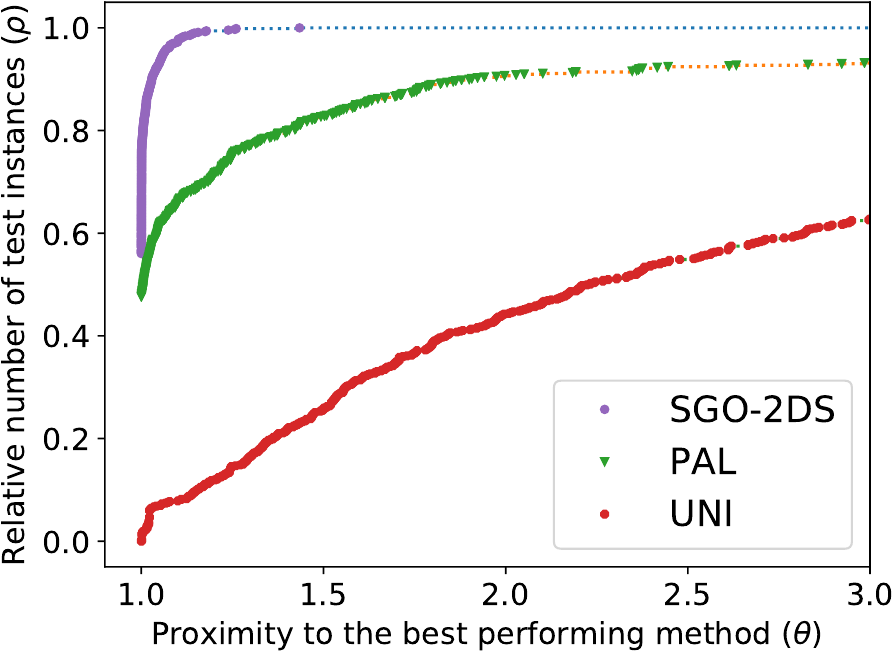}
        \caption{{\small $(32, 32)$}}
        \label{fig:asc_sym_perf_32}
    \end{subfigure}
    \caption{\small Performance profile plots of the symmetric partitioning
        methods when the graphs are reordered in ascending order of their degrees
        and only the upper triangular part is kept. The algorithms are compared wrt.~\eqref{obj_sym}.}
    \label{fig:asc_sym_perf}
\end{figure*}

\subsubsection{\texorpdfstring{$2$}{2}-dimensional Symmetric Rectilinear Partitioning}
\label{symmetric_experiments}

In this experiment, we compare \sgotwds, PAL and UNI algorithms. The objective function
tries to minimize the load of the maximum loaded partition.
As illustrated in Figure~\ref{fig:asc_sym_perf}, we see that the relative order of
the algorithms with respect to partitioning quality is \sgotwds, PAL and UNI. We
observe that the difference between \sgotwds and PAL algorithms increases
proportional to the number of parts, from $(8, 8)$ to $(32, 32)$. We also observe
that on nearly $80\%$ of the matrices, the partition quality is very close
between \sgotwds and PAL algorithms while \sgotwds outperforms the PAL algorithm
on the rest of the matrices. Therefore we claim that \sgotwds is more resistant
to the sparsity pattern of the given matrix and outputs better partitions.
As expected, the UNI algorithm performs really badly and it gives up to $3$ times worse partitions.


\subsubsection{\texorpdfstring{$3$}{3}-dimensional Rectilinear Partitioning}

\begin{figure*}[htb]
    \begin{subfigure}[b]{0.3\textwidth}
        \centering
        \includegraphics[width=\textwidth]{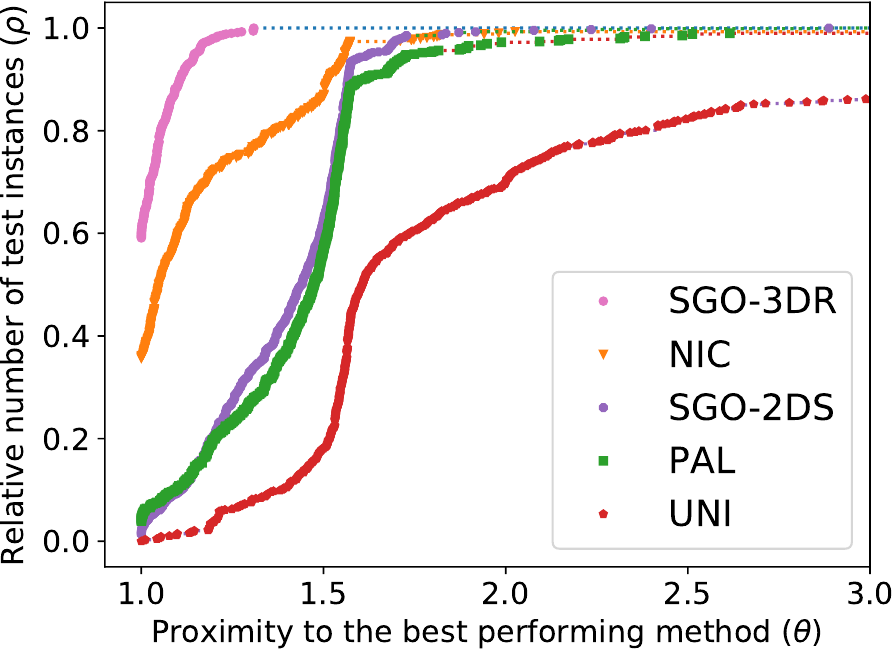}
        \caption{{\small $(8, 8)$}}
        \label{fig:spgemm_perf_8}
    \end{subfigure}
    \hfill
    \begin{subfigure}[b]{0.3\textwidth}
        \centering
        \includegraphics[width=\textwidth]{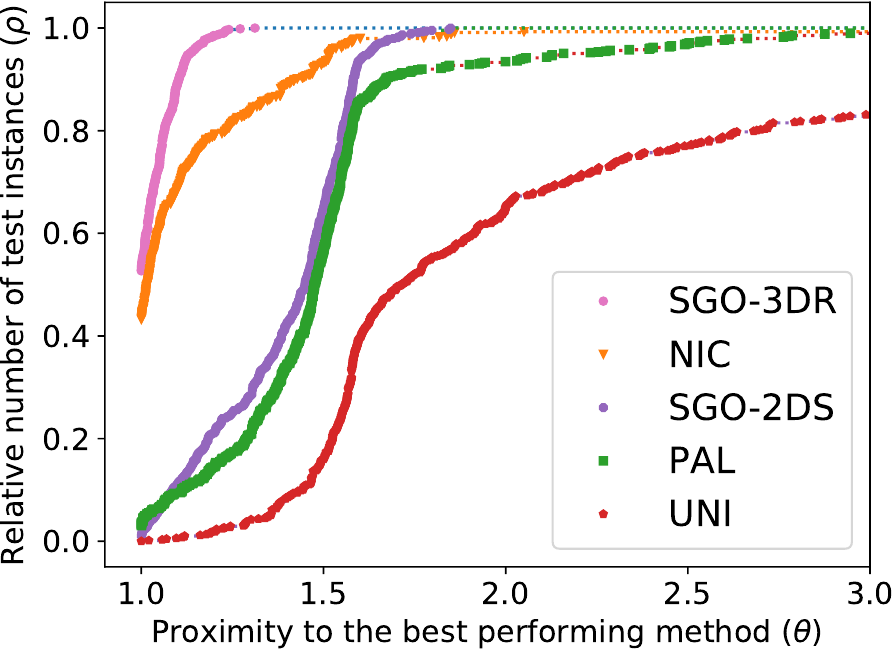}
        \caption{{\small $(16, 16)$}}
        \label{fig:spgemm_perf_16}
    \end{subfigure}
    \hfill
    \begin{subfigure}[b]{0.3\textwidth}
        \centering
        \includegraphics[width=\textwidth]{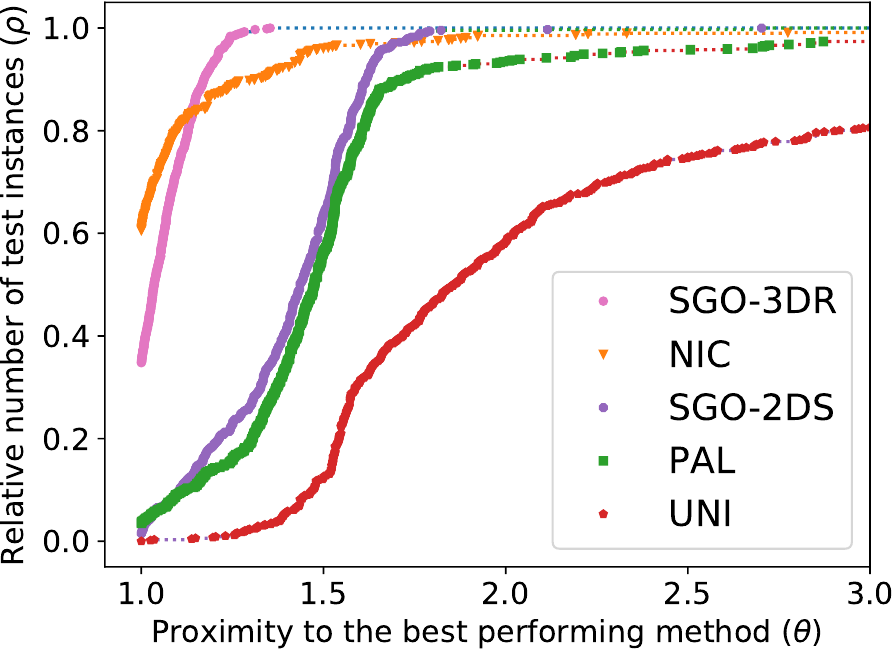}
        \caption{{\small $(32, 32)$}}
        \label{fig:spgemm_perf_32}
    \end{subfigure}
    \caption{\small Performance profile plots of the partitioning methods
        with natural reordering. The algorithms are compared wrt.~\eqref{obj_spgemm}.}
    \label{fig:spgemm_perf}
\end{figure*}

In this experiment, we compare \sgothdr, \sgotwds, NIC, PAL and UNI algorithms and
we use~\eqref{obj_spgemm} as the objective function.
In this use case, we require $3$ partition arrays $p = (p_1, p_2, p_3)$. For the symmetric
methods (\sgotwds, PAL and UNI) we use the same partition array for each dimension. However,
for the NIC algorithm, initially, we partition the first load distribution $f_A$ to get $p_1$ and $p_2$.
Then we find the optimal $p_3$ to partition $f_B$ when the partition array for the first dimension of $f_B$
is $p_2$. \sgothdr outputs $3$ partition arrays as an output. Figure~\ref{fig:spgemm_perf},
shows that the relative order of the algorithms with respect to partitioning quality is
\sgothdr, NIC, \sgotwds, PAL and UNI.

\subsubsection{\texorpdfstring{$3$}{3}-dimensional Symmetric Rectilinear
    Partitioning}

\begin{figure*}[htb]
    \begin{subfigure}[b]{0.3\textwidth}
        \centering
        \includegraphics[width=\textwidth]{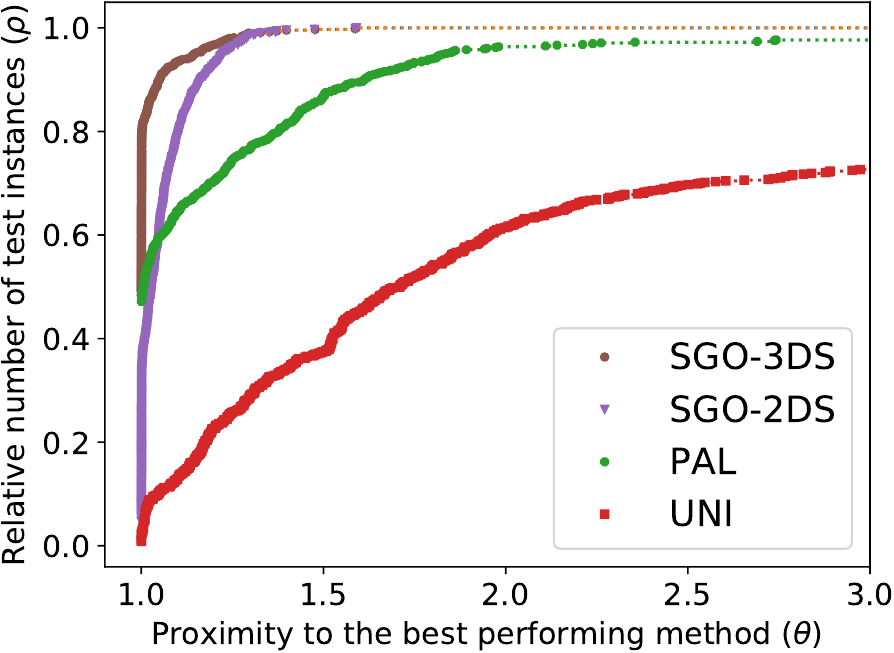}
        \caption{{\small $(8, 8)$}}
        \label{fig:sym_tri_perf_8}
    \end{subfigure}
    \hfill
    \begin{subfigure}[b]{0.3\textwidth}
        \centering
        \includegraphics[width=\textwidth]{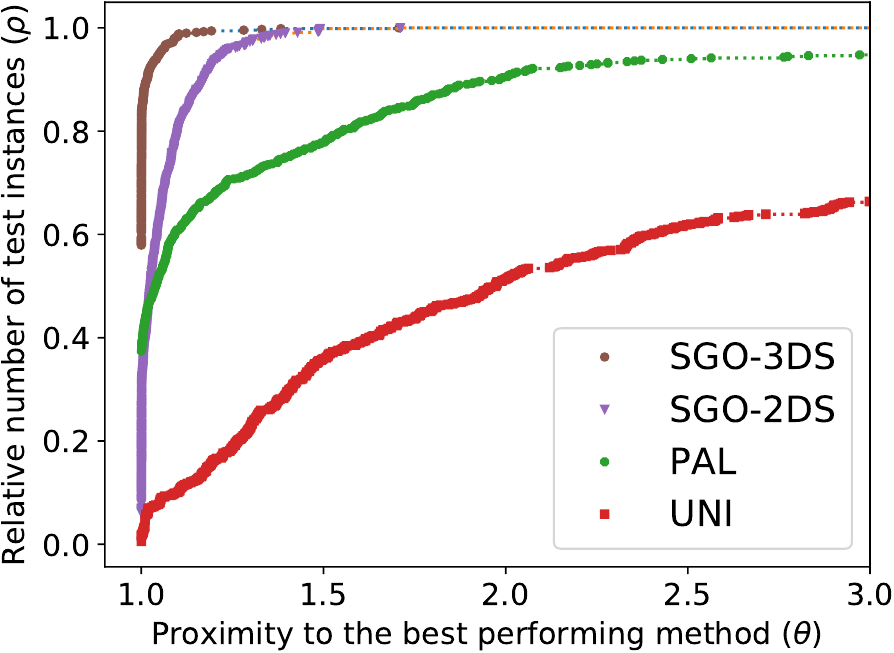}
        \caption{{\small $(16, 16)$}}
        \label{fig:sym_tri_perf_16}
    \end{subfigure}
    \hfill
    \begin{subfigure}[b]{0.3\textwidth}
        \centering
        \includegraphics[width=\textwidth]{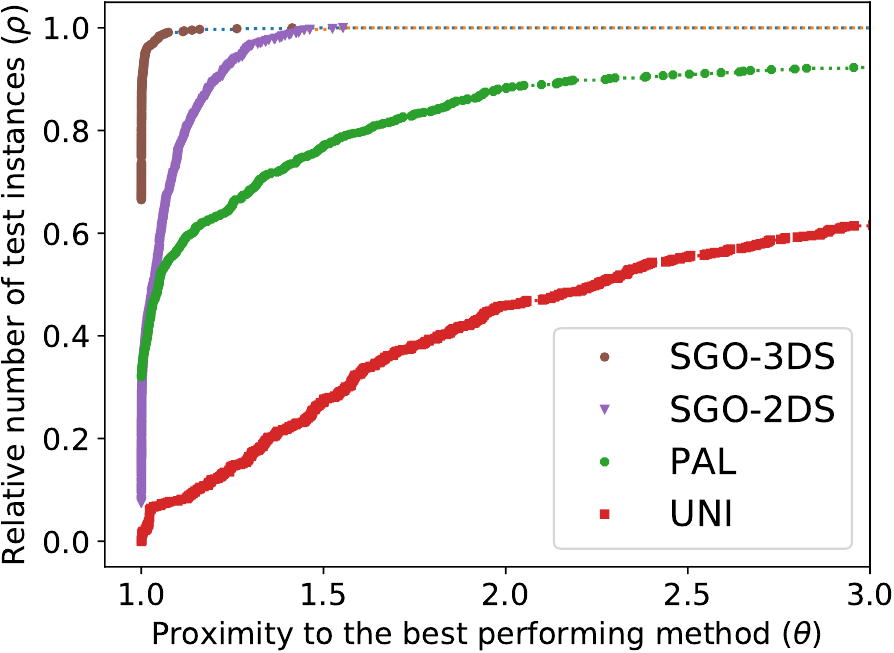}
        \caption{{\small $(32, 32)$}}
        \label{fig:sym_tri_perf_32}
    \end{subfigure}
    \caption{\small Performance profile plots of the symmetric partitioning
        methods when the graphs are reordered in ascending order of their degrees
        and only the upper triangular part is kept which is useful when doing
        triangle counting. The algorithms are compared wrt.~\eqref{obj_tri}.}
    \label{fig:sym_tri_perf}
\end{figure*}

In this experiment, we compare \sgothds, \sgotwds, PAL and UNI algorithms,
and we use~\eqref{obj_tri} as the objective function.
Figure~\ref{fig:sym_tri_perf} illustrates that the relative order of
the algorithms with respect to partitioning quality is \sgothds, \sgotwds, PAL and
UNI. We observe that the difference between \sgothds and other algorithms start to
increase as we increase the number of parts from $(8, 8)$ to $(32, 32)$. Among those
algorithms only \sgothds tries to minimize the objective function~\eqref{obj_tri}, we see
that minimizing the maximum load of a single partition also helps in most cases as
\sgotwds seems to perform relatively well. The reason for this phenomenon is that
$3$ times~\eqref{obj_sym} is an upper bound for~\eqref{obj_tri}, so optimizing
for \eqref{obj_sym} also implicitly optimizes for~\eqref{obj_sym}.

\subsection{Evaluation of the Execution Time}
\label{subsec:execution_time}

\begin{figure*}
    \begin{subfigure}[b]{0.3\textwidth}
        \centering
        \includegraphics[width=\textwidth]{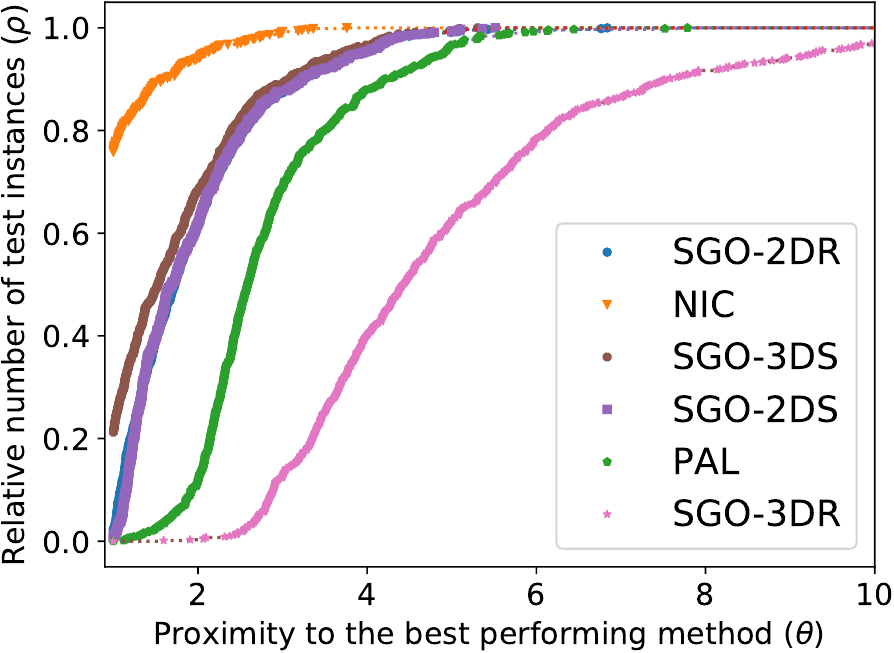}
        \caption{{\small $(8, 8)$}}
        \label{fig:exec_perf_8}
    \end{subfigure}
    \hfill
    \begin{subfigure}[b]{0.3\textwidth}
        \centering
        \includegraphics[width=\textwidth]{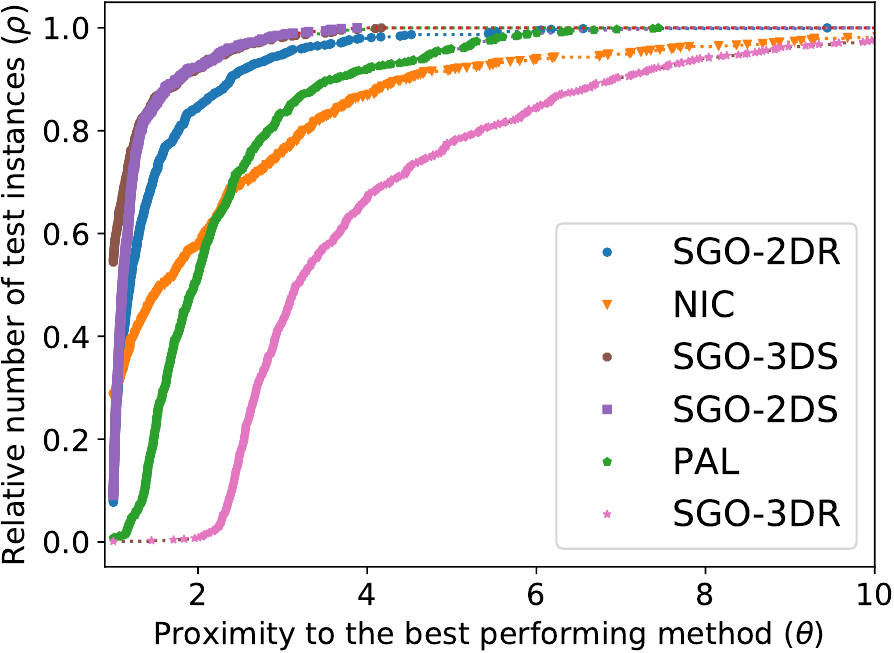}
        \caption{{\small $(16, 16)$}}
        \label{fig:exec_perf_16}
    \end{subfigure}
    \hfill
    \begin{subfigure}[b]{0.3\textwidth}
        \centering
        \includegraphics[width=\textwidth]{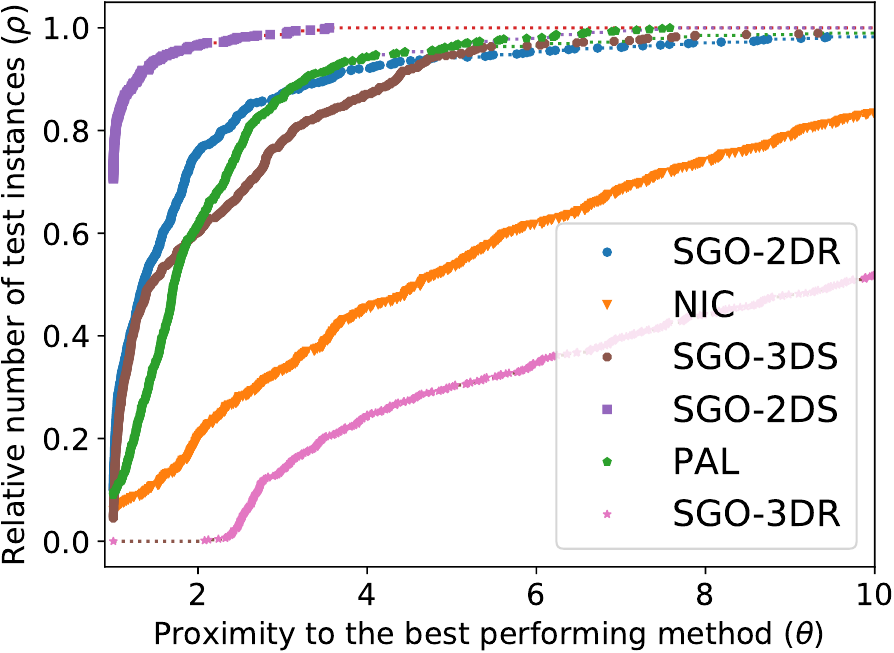}
        \caption{{\small $(32, 32)$}}
        \label{fig:exec_perf_32}
    \end{subfigure}
    \caption{\small Performance profile plots of the partitioning methods
        with natural ordering. The algorithms are compared wrt. their execution times.}
    \label{fig:exec_perf}
\end{figure*}

We would like to note that in this work our goal is to propose a novel subgradiant-based
multi-dimensional rectilinear partitioning framework. However, to achieve better performance
we enabled parallelization features using modern C++ execution policies and also we
transformed pleasingly parallelizable loops in to parallel.
All of the algorithms used for the experiments were parallelized in the same manner
including the sparse prefix sum data structure.
We include data structure construction time and partitioning time in reported execution times.
Among the RPP algorithms, we pick NIC as the baseline omitting 4APX and 2SWP as NIC is
the only algorithm that gives comparable results to \sgotwdr. The execution time of 2SWP is
around $5$ to $10$ times faster than NIC because it only does 2 iterations whereas NIC can do
upto $20$ iterations. Even though 4APX does more iterations than \sgotwdr and requires a binary
search on top as the algorithm is presented in the format of a decision procedure, it still gives worse results and the runtime of a single iteration is the same as \sgotwdr as they both
use the same sparse prefix sum data structure to query loads of tiles in each iteration.

In Figure~\ref{fig:exec_perf}, we observe that when we increase the number of parts the NIC
algorithm gives worse performance;
as we go from $k = (k_1, k_2) = (8, 8)$ to $(32, 32)$. Because the complexity of the NIC algorithm
depends on $k_1 + k_2$ linearly and other algorithms' runtimes are mostly dominated by the
sparse prefix sum data structure construction time, which doesn't depend on $k$.
In addition, we observe that different variants of \sgo are at least as fast as the PAL algorithm,
while \sgotwds being around more than 2 times faster on average. Since \sgothdr partitions two
matrices simultaneously and has to build two different sparse prefix sum data structures and
query them during each iteration, it does at least $2$ times as much work as other algorithms.
It also optimizes over $3$ partition arrays which increases the dimensionality of the searches
space by $3$ times. That is why it is much slower than other \sgo variants.

\section{Conclusion}
\label{sec:conclusion}
In this paper, we propose an efficient iterative subgradient-based method, \sgo, for the
rectilinear partitioning problem that is generalizable for arbitrary $d$ dimensions. We also show our
framework can solve the symmetric rectilinear partitioning problem, via constraints.
We propose algorithms to two variants of this problem. Finally, our experiments
on more than $600$ matrices
show that \sgo outperforms state-of-the-art algorithms in terms of partition quality
and execution time.

\section*{Acknowledgments}
This material is based upon work supported by the National Science Foundation
under Grant Number CCF-1919021.
This research was supported in part through research cyberinfrastructure resources
and services provided by the Partnership for an Advanced Computing Environment (PACE) at
the Georgia Institute of Technology, Atlanta, Georgia, USA.

\bibliographystyle{ACM-Reference-Format}
\bibliography{biblio,tdalab,tc}

\clearpage

\appendix

\section{Appendix}
\label{sec:appendix}

\subsection{Detailed results for a small subset of matrices}

In order to
provide reproducible results, we have provided normalized load imbalance
(with respect to average non-zero per part, meaning $\frac{L(f, p) k_1 k_2}{L(f)}$) and absolute execution times of the
algorithms for a selected subset of the sparse matrices.
Table~\ref{tab:properties_of_sparse_matrices} presents the detailed properties
of those matrices, total 16 of them.
\Cref{tab:li8,tab:li16,tab:li32,tab:et8,tab:et16,tab:et32} present
the normalized loads and execution times of different algorithms for $k_1 = k_2 \in \{8, 16, 32\}$.

\begin{table*}[ht]
    \centering
    \caption{\small The properties of sparse matrices.}
    \label{tab:properties_of_sparse_matrices}
    \begin{tabular}{|l|l|r|r|r|}
        \hline
        \textbf{Matrix Name} & \textbf{Matrix Origin} & \textbf{\# Rows} & \textbf{\# Nonzeros} & \textbf{Density} \\ \hline
        twitter7             & Social                 & 41,652,231       & 1,468,365,182        & 35.25            \\ \hline
        uk-2005              & Web                    & 39,459,926       & 936,364,282          & 23.73            \\ \hline
        stokes               & Semiconductor          & 11,449,534       & 349,321,980          & 30.51            \\ \hline
        kmer\_A2a            & Biological             & 170,728,176      & 180,292,586          & 1.06             \\ \hline
        nlpkkt160            & Optimization           & 8,345,601        & 118,931,856          & 14.25            \\ \hline
        com-Orkut            & Social                 & 3,072,442        & 117,185,083          & 38.14            \\ \hline
        kron\_g500-logn21    & Kronecker              & 2,097,153        & 91,042,010           & 43.41            \\ \hline
        soc-LiveJournal1     & Social                 & 4,847,572        & 68,993,773           & 14.23            \\ \hline
        Cube\_Coup\_dt6      & Structural             & 2,164,761        & 64,685,452           & 29.88            \\ \hline
        circuit5M            & Simulation             & 5,558,327        & 59,524,291           & 10.71            \\ \hline
        hollywood-2009       & Movie/Actor            & 1,139,906        & 57,515,616           & 50.46            \\ \hline
        wb-edu               & Web                    & 9,845,726        & 57,156,537           & 5.81             \\ \hline
        europe\_osm          & Road                   & 50,912,019       & 54,054,660           & 1.06             \\ \hline
        dielFilterV3real     & Electromagnetics       & 1,102,825        & 45,204,422           & 40.99            \\ \hline
        kron\_g500-logn20    & Kronecker              & 1,048,577        & 44,620,272           & 42.55            \\ \hline
        road\_usa            & Road                   & 23,947,348       & 28,854,312           & 1.20             \\ \hline
    \end{tabular}
\end{table*}

\begin{table*}[ht]
    \centering
    \caption{\small The normalized loads of different algorithms for $k_1 = k_2 = 8$ compared wrt.~\eqref{obj_rect}.}
    \label{tab:li8}
    \begin{tabular}{|l|r|r|r|r|r|r|r|}
        \hline
                          & \textbf{\sgotwdr} & \textbf{NIC} & \textbf{2SWP} & \textbf{4APX} & \textbf{\sgotwds} & \textbf{PAL} & \textbf{UNI} \\ \hline
        twitter7          & 1.78              & 1.67         & 2.32          & 1.81          & 1.85              & 1.80         & 9.35         \\ \hline
        uk-2005           & 5.06              & 5.13         & 6.29          & 7.90          & 7.65              & 7.65         & 11.47        \\ \hline
        stokes            & 2.93              & 2.75         & 3.13          & 4.27          & 4.71              & 4.79         & 5.01         \\ \hline
        kmer\_A2a         & 1.95              & 1.84         & 2.27          & 2.43          & 2.63              & 3.33         & 2.51         \\ \hline
        nlpkkt160         & 4.88              & 5.75         & 6.17          & 13.65         & 9.46              & 7.75         & 14.66        \\ \hline
        com-Orkut         & 2.78              & 2.72         & 2.80          & 2.87          & 2.87              & 2.83         & 5.95         \\ \hline
        kron\_g500-logn21 & 1.83              & 1.89         & 2.48          & 2.01          & 2.49              & 3.16         & 2.08         \\ \hline
        soc-LiveJournal1  & 2.17              & 2.13         & 2.57          & 2.68          & 2.51              & 2.51         & 15.40        \\ \hline
        Cube\_Coup\_dt6   & 4.97              & 7.66         & 6.40          & 7.70          & 7.69              & 7.69         & 7.84         \\ \hline
        circuit5M         & 2.64              & 2.79         & 3.22          & 3.64          & 2.67              & 3.19         & 16.94        \\ \hline
        hollywood-2009    & 4.20              & 4.01         & 4.75          & 6.88          & 5.78              & 5.78         & 11.05        \\ \hline
        wb-edu            & 4.79              & 5.67         & 6.05          & 7.66          & 7.63              & 7.63         & 8.78         \\ \hline
        europe\_osm       & 4.93              & 6.79         & 5.66          & 7.39          & 7.39              & 7.39         & 7.70         \\ \hline
        dielFilterV3real  & 3.06              & 3.00         & 4.88          & 4.45          & 4.66              & 4.09         & 5.60         \\ \hline
        kron\_g500-logn20 & 1.85              & 1.89         & 2.58          & 2.00          & 2.61              & 3.16         & 2.08         \\ \hline
        road\_usa         & 5.05              & 5.90         & 5.05          & 6.85          & 6.86              & 6.84         & 7.07         \\ \hline
    \end{tabular}
\end{table*}

\begin{table*}[ht]
    \centering
    \caption{\small The normalized loads of different algorithms for $k_1 = k_2 = 16$ compared wrt.~\eqref{obj_rect}.}
    \label{tab:li16}
    \begin{tabular}{|l|r|r|r|r|r|r|r|}
        \hline
                          & \textbf{\sgotwdr} & \textbf{NIC} & \textbf{2SWP} & \textbf{4APX} & \textbf{\sgotwds} & \textbf{PAL} & \textbf{UNI} \\ \hline
        twitter7          & 2.07              & 2.00         & 2.91          & 2.16          & 2.17              & 2.13         & 15.12        \\ \hline
        uk-2005           & 9.37              & 9.83         & 13.87         & 17.43         & 15.21             & 15.20        & 27.09        \\ \hline
        stokes            & 5.43              & 5.33         & 6.10          & 8.21          & 8.52              & 9.06         & 9.06         \\ \hline
        kmer\_A2a         & 2.48              & 2.46         & 3.11          & 3.25          & 3.33              & 4.03         & 4.43         \\ \hline
        nlpkkt160         & 9.25              & 7.98         & 13.34         & 27.11         & 18.06             & 15.46        & 28.83        \\ \hline
        com-Orkut         & 3.58              & 3.53         & 4.64          & 3.84          & 3.62              & 3.59         & 14.03        \\ \hline
        kron\_g500-logn21 & 1.98              & 2.26         & 2.85          & 2.01          & 3.36              & 3.54         & 2.17         \\ \hline
        soc-LiveJournal1  & 3.10              & 2.94         & 3.80          & 4.57          & 3.88              & 3.88         & 27.47        \\ \hline
        Cube\_Coup\_dt6   & 9.92              & 14.26        & 13.78         & 15.35         & 14.74             & 14.73        & 15.52        \\ \hline
        circuit5M         & 4.93              & 5.42         & 5.43          & 6.76          & 5.00              & 4.61         & 26.61        \\ \hline
        hollywood-2009    & 7.55              & 7.28         & 7.99          & 11.59         & 9.98              & 9.66         & 28.95        \\ \hline
        wb-edu            & 9.42              & 10.41        & 10.11         & 16.16         & 15.21             & 15.21        & 18.22        \\ \hline
        europe\_osm       & 8.77              & 10.68        & 9.48          & 14.64         & 14.62             & 14.61        & 15.37        \\ \hline
        dielFilterV3real  & 5.82              & 6.12         & 6.90          & 8.37          & 8.89              & 7.43         & 8.59         \\ \hline
        kron\_g500-logn20 & 2.09              & 2.25         & 3.00          & 2.07          & 3.04              & 3.54         & 2.13         \\ \hline
        road\_usa         & 8.94              & 8.94         & 11.14         & 12.32         & 12.30             & 11.48        & 13.23        \\ \hline
    \end{tabular}
\end{table*}

\begin{table*}[ht]
    \centering
    \caption{\small The normalized loads of different algorithms for $k_1 = k_2 = 32$ compared wrt.~\eqref{obj_rect}.}
    \label{tab:li32}
    \begin{tabular}{|l|r|r|r|r|r|r|r|}
        \hline
                          & \textbf{\sgotwdr} & \textbf{NIC} & \textbf{2SWP} & \textbf{4APX} & \textbf{\sgotwds} & \textbf{PAL} & \textbf{UNI} \\ \hline
        twitter7          & 2.42              & 2.47         & 3.35          & 2.88          & 2.64              & 2.63         & 26.60        \\ \hline
        uk-2005           & 18.03             & 15.84        & 19.72         & 43.52         & 30.35             & 30.29        & 57.25        \\ \hline
        stokes            & 10.55             & 10.09        & 11.71         & 16.46         & 16.45             & 16.94        & 19.10        \\ \hline
        kmer\_A2a         & 4.15              & 3.97         & 5.45          & 5.14          & 5.17              & 5.43         & 7.19         \\ \hline
        nlpkkt160         & 17.94             & 15.96        & 27.06         & 50.97         & 36.66             & 30.54        & 55.67        \\ \hline
        com-Orkut         & 4.57              & 4.38         & 6.95          & 5.26          & 4.79              & 4.80         & 23.69        \\ \hline
        kron\_g500-logn21 & 2.26              & 2.62         & 3.51          & 2.02          & 3.21              & 3.76         & 2.32         \\ \hline
        soc-LiveJournal1  & 4.80              & 4.42         & 5.72          & 11.41         & 6.28              & 6.28         & 49.99        \\ \hline
        Cube\_Coup\_dt6   & 19.85             & 21.82        & 27.85         & 27.14         & 26.93             & 26.92        & 30.65        \\ \hline
        circuit5M         & 9.55              & 8.46         & 10.44         & 18.17         & 8.66              & 8.68         & 49.37        \\ \hline
        hollywood-2009    & 12.80             & 12.48        & 14.93         & 40.53         & 17.76             & 17.74        & 93.46        \\ \hline
        wb-edu            & 18.62             & 19.23        & 19.97         & 31.67         & 30.31             & 30.30        & 40.53        \\ \hline
        europe\_osm       & 18.05             & 17.88        & 23.94         & 29.07         & 28.98             & 28.97        & 30.65        \\ \hline
        dielFilterV3real  & 11.24             & 11.17        & 12.93         & 16.14         & 16.02             & 14.17        & 17.29        \\ \hline
        kron\_g500-logn20 & 2.45              & 2.63         & 3.72          & 2.02          & 3.29              & 3.76         & 2.40         \\ \hline
        road\_usa         & 13.22             & 13.50        & 16.31         & 17.56         & 17.17             & 18.22        & 21.67        \\ \hline
    \end{tabular}
\end{table*}

\begin{table*}[ht]
    \centering
    \caption{\small The execution time in seconds of different algorithms for $k_1 = k_2 = 8$.}
    \label{tab:et8}
    \begin{tabular}{|l|r|r|r|r|r|r|}
        \hline
                          & \textbf{\sgotwdr} & \textbf{NIC} & \textbf{2SWP} & \textbf{4APX} & \textbf{\sgotwds} & \textbf{PAL} \\ \hline
        twitter7          & 127.76            & 37.38        & 739.04        & 129.64        & 127.79            & 117.90       \\ \hline
        uk-2005           & 25.17             & 15.45        & 112.86        & 40.37         & 25.17             & 29.93        \\ \hline
        stokes            & 9.98              & 3.93         & 35.65         & 15.22         & 10.00             & 11.36        \\ \hline
        kmer\_A2a         & 13.03             & 11.06        & 69.84         & 55.48         & 13.06             & 23.72        \\ \hline
        nlpkkt160         & 3.76              & 1.59         & 11.77         & 11.34         & 3.76              & 4.64         \\ \hline
        com-Orkut         & 6.00              & 2.15         & 19.15         & 11.66         & 6.02              & 6.63         \\ \hline
        kron\_g500-logn21 & 4.74              & 1.55         & 16.43         & 7.72          & 4.74              & 5.05         \\ \hline
        soc-LiveJournal1  & 3.38              & 1.54         & 12.45         & 9.08          & 3.39              & 4.44         \\ \hline
        Cube\_Coup\_dt6   & 1.62              & 0.42         & 5.98          & 5.99          & 1.62              & 1.96         \\ \hline
        circuit5M         & 1.99              & 1.09         & 5.85          & 5.91          & 1.99              & 2.64         \\ \hline
        hollywood-2009    & 1.81              & 0.80         & 6.00          & 4.98          & 1.82              & 1.99         \\ \hline
        wb-edu            & 1.88              & 1.19         & 8.49          & 7.11          & 1.88              & 2.64         \\ \hline
        europe\_osm       & 3.89              & 2.40         & 15.53         & 18.49         & 3.89              & 7.95         \\ \hline
        dielFilterV3real  & 1.27              & 0.44         & 3.93          & 3.92          & 1.27              & 1.43         \\ \hline
        kron\_g500-logn20 & 2.10              & 0.76         & 7.25          & 4.85          & 2.10              & 2.20         \\ \hline
        road\_usa         & 1.77              & 1.41         & 9.77          & 10.42         & 1.78              & 3.29         \\ \hline
    \end{tabular}
\end{table*}

\begin{table*}[ht]
    \centering
    \caption{\small The execution time in seconds of different algorithms for $k_1 = k_2 = 16$.}
    \label{tab:et16}
    \begin{tabular}{|l|r|r|r|r|r|r|}
        \hline
                          & \textbf{\sgotwdr} & \textbf{NIC} & \textbf{2SWP} & \textbf{4APX} & \textbf{\sgotwds} & \textbf{PAL} \\ \hline
        twitter7          & 127.86            & 44.22        & 753.83        & 135.45        & 127.88            & 118.16       \\ \hline
        uk-2005           & 25.22             & 33.06        & 116.03        & 45.24         & 25.19             & 29.89        \\ \hline
        stokes            & 10.05             & 8.21         & 32.64         & 21.05         & 10.02             & 11.45        \\ \hline
        kmer\_A2a         & 13.21             & 33.33        & 79.05         & 56.92         & 13.10             & 24.34        \\ \hline
        nlpkkt160         & 3.77              & 3.69         & 12.05         & 13.23         & 3.77              & 4.71         \\ \hline
        com-Orkut         & 6.04              & 2.02         & 18.89         & 13.51         & 6.07              & 6.72         \\ \hline
        kron\_g500-logn21 & 4.80              & 2.08         & 16.69         & 9.19          & 4.81              & 5.13         \\ \hline
        soc-LiveJournal1  & 3.43              & 3.43         & 12.28         & 11.44         & 3.43              & 4.53         \\ \hline
        Cube\_Coup\_dt6   & 1.63              & 0.71         & 6.10          & 6.48          & 1.63              & 2.04         \\ \hline
        circuit5M         & 2.02              & 2.00         & 6.06          & 7.26          & 2.02              & 2.67         \\ \hline
        hollywood-2009    & 1.84              & 1.77         & 6.37          & 8.39          & 1.84              & 2.08         \\ \hline
        wb-edu            & 1.89              & 3.08         & 9.07          & 11.99         & 1.88              & 2.76         \\ \hline
        europe\_osm       & 3.90              & 6.95         & 17.70         & 19.99         & 3.90              & 7.94         \\ \hline
        dielFilterV3real  & 1.30              & 0.65         & 4.04          & 6.55          & 1.28              & 1.50         \\ \hline
        kron\_g500-logn20 & 2.17              & 1.26         & 7.11          & 5.63          & 2.15              & 2.28         \\ \hline
        road\_usa         & 1.78              & 5.04         & 10.52         & 13.62         & 1.82              & 3.42         \\ \hline
    \end{tabular}
\end{table*}

\begin{table*}[ht]
    \centering
    \caption{\small The execution time in seconds of different algorithms for $k_1 = k_2 = 32$.}
    \label{tab:et32}
    \begin{tabular}{|l|r|r|r|r|r|r|}
        \hline
                          & \textbf{\sgotwdr} & \textbf{NIC} & \textbf{2SWP} & \textbf{4APX} & \textbf{\sgotwds} & \textbf{PAL} \\ \hline
        twitter7          & 128.47            & 138.33       & 248.65        & 152.22        & 128.34            & 118.50       \\ \hline
        uk-2005           & 25.53             & 67.08        & 125.55        & 51.70         & 25.30             & 30.31        \\ \hline
        stokes            & 10.41             & 20.48        & 33.74         & 26.21         & 10.15             & 11.72        \\ \hline
        kmer\_A2a         & 13.63             & 54.22        & 82.41         & 65.21         & 13.47             & 25.05        \\ \hline
        nlpkkt160         & 3.80              & 10.03        & 12.64         & 24.14         & 3.79              & 4.93         \\ \hline
        com-Orkut         & 6.35              & 3.96         & 19.13         & 16.63         & 6.27              & 6.94         \\ \hline
        kron\_g500-logn21 & 5.35              & 3.47         & 16.83         & 12.71         & 5.17              & 5.32         \\ \hline
        soc-LiveJournal1  & 3.78              & 9.20         & 12.65         & 15.24         & 3.59              & 4.76         \\ \hline
        Cube\_Coup\_dt6   & 1.64              & 4.20         & 6.37          & 12.73         & 1.64              & 2.24         \\ \hline
        circuit5M         & 2.16              & 3.55         & 7.02          & 9.11          & 2.11              & 2.86         \\ \hline
        hollywood-2009    & 2.04              & 4.16         & 6.38          & 10.82         & 1.88              & 2.28         \\ \hline
        wb-edu            & 1.97              & 9.62         & 9.66          & 17.50         & 1.91              & 3.01         \\ \hline
        europe\_osm       & 3.98              & 35.89        & 20.03         & 26.20         & 3.92              & 8.46         \\ \hline
        dielFilterV3real  & 1.41              & 2.96         & 4.26          & 10.90         & 1.37              & 1.67         \\ \hline
        kron\_g500-logn20 & 2.73              & 1.82         & 7.13          & 7.39          & 2.53              & 2.44         \\ \hline
        road\_usa         & 1.86              & 12.40        & 11.45         & 14.81         & 1.95              & 3.59         \\ \hline
    \end{tabular}
\end{table*}

\end{document}